\documentstyle[aps,prl,multicol]{revtex}
\include{psfig}
\input epsf
\begin{document}
\author{E. Zaccarelli$^1$, G. Foffi$^{1,2}$, K.A. Dawson$^2$, S.V.
  Buldyrev$^{3}$, F. Sciortino$^1$ and P. Tartaglia$^1$} \address{
  $^1$ Dipartimento di Fisica, Istituto Nazionale di Fisica della
  Materia, and INFM Center for Statistical Mechanics \\ and
  Complexity, Universit\`{a} di Roma La Sapienza, P.le A. Moro 5,
  I-00185 Rome, Italy\\ $^2$ Irish Centre for Colloid Science and
  Biomaterials, Department of Chemistry,\\ University College Dublin,
  Belfield, Dublin 4, Ireland\\ $^3$ Center for Polymer Studies and
  Department of Physics, Boston University, Boston, MA 02215, USA.  }
\title{ Confirmation of Anomalous Dynamical Arrest in attractive
  colloids: a molecular dynamics study.}
\date{\today}
\maketitle
\begin{abstract}
Previous theoretical, along with early simulation and experimental,
studies have indicated that particles with a short-ranged attraction
exhibit a range of new dynamical arrest phenomena. These include very
pronounced reentrance in the dynamical arrest curve, a logarithmic
singularity in the density correlation functions, and the existence of
`attractive' and `repulsive' glasses. Here we carry out extensive
molecular dynamics calculations on dense systems interacting via a
square-well potential.  This is one of the simplest systems with the
required properties, and may be regarded as canonical for interpreting
the phase diagram, and now also the dynamical arrest. We confirm the
theoretical predictions for re-entrance, logarithmic singularity, and
give the first direct evidence of the coexistence, independent of
theory, of the two coexisting glasses.  We now regard the previous
predictions of these phenomena as having been established.
\end{abstract}    


\begin{multicols}{2}
\section{Introduction}
%%%Attractive Colloidal systems
Recently, there has emerged a series of remarkable results involving
the dynamical arrest of particles with interaction potentials that are
short compared to the size of the repulsive core
\cite{bergenholtz99,fabbian99,foffi00,dawson00,zaccarelli01,foffi02,merida,puertas02,science02}.
The most important predictions are as follows \cite{cocisreview}. The
curve at which the fluid phase is arrested is reentrant in the regime
where attractions and repulsions compete. In the vicinity of this
reentrance, and in the arrested regime, there exists two distinct
arrested states that may, under some conditions, coexist. These two
states differ in their long-term dynamics.  This coexistence
terminates smoothly, the dynamics of the two arrested states becoming
identical at a particular point at which distinctive dynamics is
expected. It is important to note that all these phenomena are
predicted to occur irrespective of the shape of the potential
(explicit results exist for square well or hard-core Yukawa
interaction models
\cite{dawson00,zaccarelli01,foffi02,merida}). The essential feature is
only that the range of the attraction in comparison to the hard core
be extremely small.
  
The results described above were first deduced
from the mode-coupling-theory (MCT) \cite{bengt,goetze91}, and we may
note that the conclusions do not depend on details of the
approximation for input static structure, Percus-Yevick, Mean
Spherical Approximation and Self-Consistent Ornstein-Zernike
Approximation, all leading to essentially the same picture. Within MCT
the merging of the two arrested states is predicted to be a higher
order dynamical singularity, and density correlations in its vicinity
are expected to obey a highly distinctive logarithmic relaxation
\cite{dawson00,sperl02}, in contrast to the conventional ergodic-non
ergodic transition where a two-step process takes place
\cite{goetze91,goetze99}.
 
More recently this decay has been observed also in several
experimental systems \cite{bartsch,mallamace,science02}, thus 
giving the first evidence that these singularities do exist in 
nature. 

Two recent numerical works \cite{puertas02,RComm02} have focused on the
new dynamical features characterizing attractive colloidal systems. In
Ref. \cite{puertas02}, colloidal interactions were modeled with a
potential chosen in such a way to avoid undesired effects such as
liquid-gas separation at low densities. Polydispersity was included to
prevent crystallization at high densities. In Ref.~\cite{RComm02}
Molecular Dynamics simulations of a square well (SW) system with very
narrow range of attraction have been performed.  In
both cases, the results have been shown to be in excellent agreement
with MCT predictions.  The focus on the simple square well potential,
makes direct contact both with the available theoretical results and
with experimental systems, being the interactions completely
controllable and not dependent on external parameters.  For the SW
model, theoretical solution is also available.
Indeed, the equilibrium properties of square-well potential have been
studied for many years, and in many ways is regarded as the simplest
physical model that exhibits all of the essential features of the
short-ranged systems \cite{alder,noro,elliot,vega,chang,rascon}.
Besides this, it is the model for which the theory of the dynamical
arrest transition described above has been
developed \cite{dawson00,zaccarelli01} and for which, therefore,
detailed comparison may be made between theory and simulation. It may
therefore be regarded as canonical in this arena of study.
 
In Ref.~\cite{RComm02}, we examined a one-component square 
well system. We found that, no matter the sharp 
intervening of crystallization which effectively prevented the system
to approach very closely the glass transition, it was possible to have
a clear picture of the reentrant shape of the glass curve in the
temperature-density plane. This was achieved by plotting
iso-diffusivity curves, and examining their trends when approaching
the limit $D\rightarrow 0$. The shape of the liquid-glass line found
was in good agreement with the previous MCT \cite{dawson00}
predictions.

In this paper, we report an extensive numerical study of a
two-component binary mixture with interactions modeled by a very
narrow SW potential.  This mixture appears to be a logical
extension of the one-component system for which crystallization is
effectively avoided, by means of the geometrical rearrangements
allowed by asymmetry in diameters of different particle species. The
choice of a two-component system makes possible to extend the range of
iso-diffusivity curves of almost $4$ decades, as well as the range of
studied packing fractions from approximately $0.57$ for one-component
to $0.62$ for binary systems.  One notes that at that highest density, the 
reentrance is so pronounced that an equivalent hard sphere system 
would be sufficiently dense to approach the 'random-close-packed' 
limit, though the definition of that concept has its own 
limitations \cite{torquato}.

The extension to a binary system does
not only make more evident the reentrant behaviour of the glass
transition curve, but also allows a deeper study of the dynamics of
the system. By this we mean that it turns out to be possible to reach
very closely the ideal glass line, intended as the $D\rightarrow0$
locus of the $(\phi,T)$-plane, and thus to feel quite clearly the
presence of higher order MCT singularities. We know from the
theoretical work that the $3\%$-case of the square well model
possesses indeed an $A_3$-singularity \cite{dawson00,zaccarelli01},
corresponding to the end-point of the repulsive-attractive glass-glass
transition. Though, of course, this singularity lies inside the glassy
region, its presence is signaled by the characteristic logarithmic
decay of density correlators mentioned above, already in the liquid
region, when going sufficiently close to the neighbouring glass
boundary.

We report in this paper a characteristic behaviour of the density
correlators, near the ideal glass transition, which combines features
of the typical $A_2$ singularity, i.e. simply ergodic to non-ergodic
transition with two-step power-law relaxation, and of higher order
singularities associated with logarithmic decay. This produces, in a
certain region of the $(\phi,T)$-plane, a new behaviour which cannot
be described by the asymptotic predictions for either of the two
behaviours, deriving by a competition of the two types of MCT
solutions. The same kind of interesting result is found in the mean
square displacement (MSD).  This subtle interplay between different
singularities is present also in theory, and its manifestation in
simulation may be regarded as support for the theory, rather than an
inconvenience.  This will allow us to identify and localize quite
clearly a genuine MCT higher order singularity in a realistic model.

\section{Simulation and Theory}
We study binary mixtures of SW spheres. In particular, we focus on
samples of a $50\%-50\%$ mixture of $N=700$ spheres in a cubic box,
with diameter ratio between the two species equal to $1.2$. Thus, the
smaller particles ($B$-type) diameter is $\sigma_B=1$, and both $A$
and $B$ particles have unit mass.  Both species interact with a square
well potential with ratio between potential range and particle
diameter equal to $3\%$. This corresponds to one of the cases studied
theoretically within MCT, that clearly possesses all the main
phenomena \cite{dawson00,zaccarelli01,chen01}.  The $3\%$-ratio has
been chosen also for interactions between particles of different
species, i.e. we consider
\begin{eqnarray}
V_{ij}(r)&=&\infty \ \ r_{ij}<\sigma_{ij} \nonumber\\
V_{ij}(r)&=&-u_0 \ \ \sigma_{ij}<r_{ij}<\sigma_{ij}+\Delta_{ij}\nonumber\\
V_{ij}(r)&=&0 \ \ r_{ij}>\sigma_{ij}+\Delta_{ij}
\end{eqnarray}
with $\epsilon_{ij}=\Delta_{ij}/(\sigma_{ij}+\Delta_{ij})=0.03$,
$i,j=A,B$ and we use the conventional notation for which, for example,
$\sigma_{AA}=\sigma_A$ and $\sigma_{AB}=(\sigma_A+\sigma_B)/2$.
Temperature $T$ is measured in units of energy, i.e. $k_B=1$ and thus,
for example, $T=1$ corresponds to the system having a thermal energy
equal to the well-depth, while the packing fraction is defined as
$\phi=(\rho_A\sigma_A^3+\rho_B\sigma_B^3)\cdot \pi/6$, where
$\rho_i=N_i/L^3$, $L$ being the box size and $N_i$ the number of
particles for each species.

Initial configurations for each density were chosen at random.
Particles were separated in successive steps, with more particular
care the higher the density of interest, to implement the hard core
repulsion. When separation was ensured, the attractive well was added.
To reach the temperature of study, the configuration so prepared was
then left to evolve with a thermostat of constant thermal coefficient
for a period of time sufficient to equilibrate at that temperature. We
estimate the equilibration time as the time at which the density
correlation function of the slowest collective mode (i.e. at the
structure factor peak) has decayed to zero.  After this equilibration,
the configuration was left to run at constant energy for a time
dependent on the slowness of the dynamics, for a time covering at
least $10$ equilibration times.

Simulation time for each species is measured in units of
$\sigma_i\cdot(m/u_0)^{1/2}$.  Standard MD algorithm has been
implemented for particles interacting with SW
potentials\cite{rapaport}. Between collisions, particles move along
straight lines with constant velocities. When the distance between the
particles becomes equal to the distance for which $V(r)$ has a
discontinuity, the velocities of the interacting particles
instantaneously change.  The algorithm calculates the shortest
collision time in the system and propagate the trajectory from one
collision to the next one.  Calculations of the next collision time
are optimized by dividing the system in small subsystems, so that
collision time are computed only between particles in the neighboring
subsystems.

We studied eight isothermal cuts of the phase diagrams, with
temperature varying between $2.0$ and $0.3$ in the large packing
fraction region, i.e. $\phi>0.5$.  In addition, we examined the hard
spheres case, where no attractive interactions are present. For each
considered configuration, we first studied the thermal history to
check that, effectively, it maintains itself at the required
temperature within fluctuations and that the total energy remains
constant.

Of each studied configuration, we considered the time-dependent
density correlation functions for different $q$-vectors to make direct
comparison with the behaviour predicted by MCT. The correlators are
defined as,
\begin{equation}
\phi_{ij}(q,t)=\frac{S_{ij}(q,t)}{S_{ij}(q)}=
\frac{\langle \rho_i(-q,t)\rho_j(q,0)\rangle}
{\langle \rho_i(-q,0)\rho_j(q,0)\rangle}
\end{equation}
where $\rho({\bf q},t)=\sum_l \exp{(i {\bf q}\cdot {\bf r}_l(t))}$ are
the density variables for each species and $S_{ij}(q)$ are the 
partial static structure factors of the system. The   
$\phi_{ij}(q,t)$ are the fundamental 
quantities of interest in the Mode Coupling theory, which consists of 
writing a set of generalised Langevin equations, that can be closed 
within certain approximations 
\cite{bengt,goetze91,lettermct,longmct}.  

The correlators $\phi_{ij}(q,t)$ have been calculated averaging over
several independent configurations and over up to 100 different
wave-vectors with the same modulus, to obtain a good statistical
sample.  Of course, these quantities show an interesting behaviour
where the dynamics of the system gets slower, i.e. in the supercooled
regime. Indeed, the relaxation process starts to show a separation in
two time-scales, which originates the typical two-step relaxation
scenario near the (ideal) glass transition.  A first relaxation
process, the so called $\beta$-relaxation, occurs at short times, due
to particles exploring the cage formed by their neighbours, while a
second process, the $\alpha$-relaxation, that accounts for the
restoration of the ergodicity due to breaking of the cages, occurs at
larger and larger time scales, due to the slowing down of the
dynamics, forming a longer and longer plateau region. At the ideal
glass transition, the time of the $\alpha$-relaxation diverges, and
the correlators do not relax anymore, thus remaining at this plateau
value. This is defined as the non-ergodicity parameter
$f_{ij}(q)=\lim_{t\rightarrow \infty}\phi_{ij}(q,t)$, which jumps
discontinuously from zero to a finite (critical) value $f^c_{ij}(q)$,
signaling the occurrence of an ergodic (fluid) to non-ergodic (glass)
transition.

The two-step relaxation is well described by MCT, through an
asymptotic study of the correlators near the ideal glass solutions.
The approach to the plateau is described by a power law, regulated by the
exponent $a$, i.e.
\begin{equation}
\phi_q(t)- f^c_q \sim h^{(1)}_q (t/t_0)^{-a}+h^{(2)}_q(t/t_0)^{-2a}
\end{equation}
with $t_0$ the microscopic time, while the departure from the plateau,
i.e. the start of the $\alpha$-process, is expressed in terms of
another power law, regulated by the exponent $b$,
\begin{equation}
\phi_q(t)- f^c_q \sim h^{(1)}_q (t/\tau)^{b}+h^{(2)}_q(t/\tau)^{2b}
\label{b-law}
\end{equation}
with $\tau$ the characteristic time of the relaxation.  The exponents
$a$ and $b$ are related to each other with an algebraic relation, and
are independent of the particular $q$-vector considered, $h^{(1)}_q$
and $h^{(2)}_q$ are called critical amplitudes \cite{goetze91}.

On the other hand, the $\alpha-$relaxation process can be also well
described by a stretched exponential, i.e.
\begin{equation}
\phi_q(t)=A_q\exp{[-(t/\tau_q)^{\beta_q}]}
\label{stretched}
\end{equation}
where the amplitude $A_q$ determines the plateau value, 
and the exponent $\beta_q$ is always less than $1$.

We can thus fit the correlators, for each $q$-value considered, both
in terms of the MCT prediction and of the stretched exponential, and
find an estimate of the non-ergodicity factor $f_q$ as a function of
$q$.  The shape of this quantity can be indicative of the formation of
different types of glasses, either attractive or  repulsive dominated
\cite{bergenholtz99,fabbian99,dawson00,zaccarelli01}.

However, what we have described so far is typical for $A_2$
singularities. These correspond to the simplest non-trivial solutions
for the non-ergodicity parameter MCT equations, and for example in the
hard sphere model only this type of singularity can arise. This is due
to the fact that the only control parameter of the model is the
packing fraction. When the number of control parameters increases,
higher order singularities may occur. For a square well model it was
shown that singularities of types $A_3$ and $A_4$ are present within
the theory \cite{dawson00} when the width of the well becomes much
smaller than the hard core radius. In the proximity of such
singularities, the asymptotic behaviour for the density correlators is
different from the one we have seen so far \cite{dawson00,sperl02}.
In particular, we have for the leading contribution a logarithmic
behaviour,
\begin{equation}
\phi_q(t)- f^c_q \sim - C_q \ ln(t/\tau)
\label{log}
\end{equation}

Another main focus of our study was to evaluate the MSD
$\langle r^2(t)\rangle$ of particles with respect to
their initial positions. We considered a configuration to have
sufficiently evolved when particles on average have traveled further
than a few diameters.  

Typically, the behaviour of the MSD at short times, follows the simple
law $\langle r^2(t)
\rangle \sim t^2$, which accounts for the ballistic motion of the
particles, i.e. particles move freely without collisions. At later
times, particles start to feel the presence of each other and there is
crossover to the diffusive regime, i.e. $\langle r^2(t) \rangle \sim
t$. The proportionality constant of this relation defines the
diffusivity of the system, via the celebrated Einstein relation \cite{hansen},
\begin{equation}
\lim_{t\rightarrow \infty}\frac{\langle r^2(t)\rangle}{t}\simeq 6D
\end{equation}
Thus, from evaluating the long time limit of the MSD, we determine the
diffusion coefficient $D$ for each state point.

Though this general behaviour is preserved, when the dynamics becomes
slower, a transient region between short and late times emerges. Of
course, the duration of this transient region increases the slower the
dynamics. This phenomenon is the correspondent behaviour for the MSD
of the separation of the two time scales that we have seen in the
correlation functions. It also reflects the formation of cages in
which particles get trapped, so that diffusion becomes more difficult.
The crossover region consists generally in the development of a
plateau also for the MSD. At the ideal glass transition, the
$\alpha-$relaxation time would diverge, and diffusion from the plateau
would not occur even at infinite times, in complete analogy with the
correlators behaviour.  Thus, the height of the plateau represents the
localization length of particles in the arrested state, i.e. the size
of the cages of the glass.  Of course, in simulations, only finite
times can be explored and the position of the ideal glass transition
can be only extrapolated by data.

As for the correlators, the presence of higher order singularities may
affect the plateau region of MSD, giving rise to peculiar new
behaviour.

MCT predicts a power law decrease for the diffusivity 
on approaching the ideal glass transition. Along an isotherm  
\begin{equation}
\label{MCT-diffu}
D\sim |\phi-\phi_c|^\gamma
\end{equation} 
where $\phi_c$ is the value of the packing fraction at the transition
(`critical' value), i.e. the value where the diffusivity would drop to
zero for the considered temperature.  The exponent $\gamma$ is
completely determined by the theory in terms of the exponents $a$ and
$b$, via the simple relation $\gamma=1/(2a)+1/(2b)$
\cite{goetze91}. It is also related to the so-called exponent
parameter $\lambda$ by an analytical relation. This parameter is
crucial in determining the presence of higher order singularities
\cite{dawson00}, in particular it tends to the value $\lambda=1$ at an
$A_3$-point, while for a simple $A_2$ singularity is always less than
$1$.  

So, in principle, from fitting the diffusivity behaviour at constant
temperature with Eq.~\ref{MCT-diffu}, one can determine the exponent
$\gamma$, and from this, also $a$ and $b$. From a fitting procedure
based on Eq.~\ref{b-law}, the non-ergodicity parameter can be
calculated consistently.  However, close to higher order
singularities, we do not expect that this behaviour is generally
preserved, as the logarithmic behaviour in the correlators and the
transient region in the MSD intervene. Indeed, in these conditions,
the exponent $b$ tends to zero (thus originating the logarithmic
behaviour) and, from the relation between $b$ and $\gamma$, one can
see that $\gamma$ would go to $\infty$, and Eq.~\ref{MCT-diffu} cannot
then describe the arrest. Thus, the region of validity of asymptotic
predictions may shrink significantly close to higher order
singularities.

\section{Results: The Overall Picture}
\begin{figure}
%\hbox to\hsize{\epsfxsize=0.8\hsize\hfil\epsfbox{fig1.eps}
%\hfil}
%\vspace*{0.5cm}
\centerline{\psfig{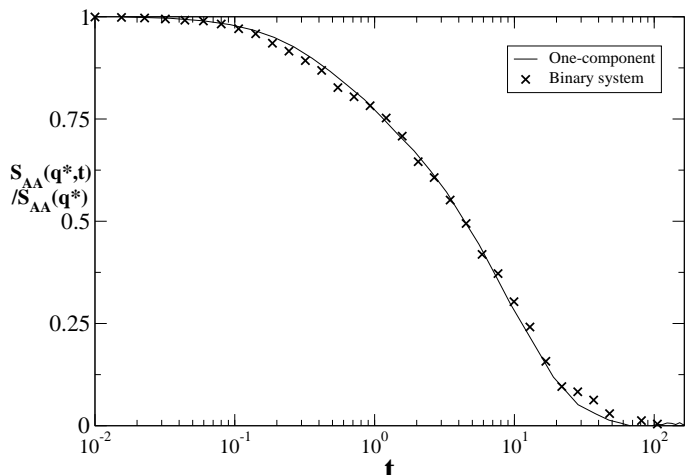}
           } 
\caption{
  Comparison between the density-density correlation function
  $S_{AA}(q^*,t)/S_{AA}(q^*)$ in the binary and the monodisperse case
  at $\phi=0.57$ and $T=0.75$, corresponding to the most reentrant
  point found in the mono-disperse case, before crystallization
  intervenes. The wave-vector chosen corresponds for both cases to
  $q^*=2\pi/\sigma_A.$}
  %Comparison between the density-density correlation function
%  $S_{AA}(q^*,t)/S_{AA}(q*)$ in the binary and the monodisperse case
%  at $\phi=0.57$ and $T=0.75$, corresponding to the most reentrant
%  point found in the mono-disperse case, before crystallization
%  intervenes. The wave-vector chosen corresponds for both cases to
%  $q^*\sigma=2\pi$.}
\vspace*{0.5cm}
\label{fig:monobina}
\end{figure}

The considered mixture represents a natural extension of the
mono-disperse system studied so far \cite{RComm02}. Indeed, the small
asymmetry in diameter does not produce significant change in the
dynamics of the two cases, and, on the contrary, allows to reach much
larger packing fractions with no sign of crystalline order. The first
of the two sentences can be explained by looking at Figure
\ref{fig:monobina}. Here, we compare the behaviour of density
correlators at a correspondent point on the control parameter space of
the system. For the mono-disperse case we are almost at the most
reentrant point before crystallization takes place, i.e. $\phi=0.57$
and $T=0.75$. We can clearly see that, nonetheless, dynamics does not
appear to be particularly slowed down.  To make the comparison with
the binary case, we are considering the total density correlation
function for the species $1$ at the $q$-vector corresponding
approximately to the first peak of the static structure factor, and
quantities have been rescaled in order to compare particles with equal
diameters.  It is evident that the dynamical behaviours are very
close, thus in this sense, we can think of doing an extension of the
one-component work.

In the following, we will focus on properties of particles of type $A$.
Thus all quantities reported without label, will refer to them. 
This choice derives from the fact that we do not
expect substantial differences in the behaviour of the two species,
due to the small amount of asymmetry in their sizes.

We start by comparing the $T$ and $\phi$ dependence of the diffusion
coefficient.  The diffusion coefficients can be normalised with
respect to the factor $D_0^i=\sigma_i \sqrt{T/m}$, which takes into
account the $T$-dependence of the microscopic time. This ensures that
the difference in the average velocities due to the temperature is
eliminated, and the diffusion can be considered to be comparable
between different temperatures.  

\begin{figure}
\centerline{\psfig{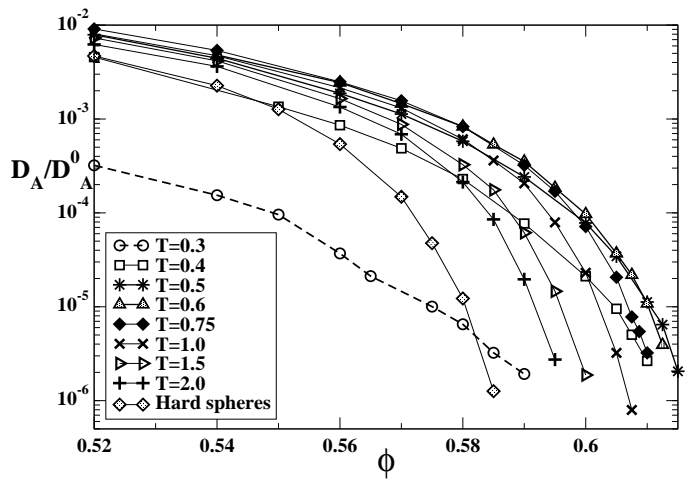}
           }
\vspace*{0.5cm}
\caption{
  Normalised diffusion costant $D/D_0$, with $D_0=\sigma\sqrt{T/m}$ in
  function of packing fraction $\phi$, along each studied isotherms
  between $T=0.3$ and $T=2.0$. The normalisation factor takes into
  account the difference due to different inital velocities, and
  ensures the common low density limit (see \protect\cite{RComm02}). }
\label{fig:diffusivity}
\end{figure}

A plot of the $D_{A}/D_0^A$ in function of packing fraction, along the
considered isotherms, is shown in Fig.~\ref{fig:diffusivity}.  In the
present work, we focus our attention mainly on the high densities
regime, i.e. $\phi>0.5$. The behaviour of the diffusivities presents
many similarities with the case of a mono-disperse sample of SW
spheres \cite{RComm02}. However, striking novel features appear due to
the fact that, for the chosen binary system, it is possible to reach
diffusivities 4 orders  of magnitude smaller than in the
mono-disperse case, as well as much larger packing fractions with no
sign of crystallization.

We present results of normalised diffusivities varying roughly between
$10^{-2}$ and $10^{-6}$, while the mono-disperse system could only
reach values up to slightly above $10^{-2}$, due to the intervening of
crystallization \cite{RComm02}. We remark that the lowest diffusivity
values were imposed by computational times, and not as in the
mono-disperse by crystallization. Also, the attractive binary system
is able to occupy effectively a larger amount of available volume,
thus reaching liquid states up to a packing of about $0.62$.  On the
other hand, the hard spheres case reaches comparable values of
diffusivities at a packing fraction of about $\phi=0.585$ (see
Fig.~\ref{fig:isodiff}), a value close to the one experimentally
established for the one-component hard sphere case
\cite{vanmegen}.
 
Examining the figure, it is evident that the behaviour of the
diffusivity is driven by two competing mechanisms. Upon decreasing
temperature starting from the highest value, the presence of the
repulsive core, initially dominant, enters in competition with the
attractive interactions. This is manifested in the diffusion getting
larger, at the correspondent packing fraction, by decreasing
temperature. In other words, the system reaches the same diffusivity
at larger and larger packing fractions. This is due mainly to
geometrical rearrangement of particles, as the temperature is lowered,
i.e. particles tend to get closer and, consequently, more free space
for diffusion gets available to the system. However, when the
temperature becomes small enough, i.e. effectively lower than the
energy scale of the square well, attractions become dominant and thus,
diffusion becomes slower again because particles tend to remain within
each other shell of attraction.
 
We know from theoretical calculations within MCT that this phenomenon
is typical of very narrow attractive potentials, both for square well
interactions \cite{dawson00} and for hard-core Yukawa \cite{foffi02}.
Indeed, the $3\%$-choice for the range of the attractive well in our
simulation corresponds to ensure that competition between attraction
and repulsion is particularly sharp. This can be explained in terms of
cages, i.e. when the attractive range is not small enough, which means
few percents of diameter, there is not much difference between cages
formed by neighbouring particles at high or low temperatures. On the
other hand, a very localised attraction can effectively change the
shape of the cages, by sticking particles within the well-distance
$\Delta$. This produces the larger diffusions observed at intermediate
temperatures, when the two mechanisms almost balance each other.
Similarly, extremely short-ranged attractions produce a solid-solid
iso-structural transition, between an attractive-dominated and a
repulsive-dominated crystal \cite{iso-frenkel}. This has been
correlated with the glass-glass transition predicted by MCT in a
recent work \cite{foffi02}. Of course, it would be interesting to
extend the simulations to different values of the range of the
potential to confirm the width dependence of the anomalous behaviour.

Observing more carefully Figure~\ref{fig:diffusivity}, the two
mechanisms of diffusion produce two different trends of behaviour for
the plotted curves. Indeed, for $T>0.6$ the curves present a quite
dramatic decrease of diffusivity, while for smaller
temperatures the same decrease of about 4 orders of magnitude occurs
on a much wider range of packing fractions (for example comparing $T=
2.0$ and $T=0.4$ such range almost doubles).
 
\begin{figure}
\centerline{\psfig{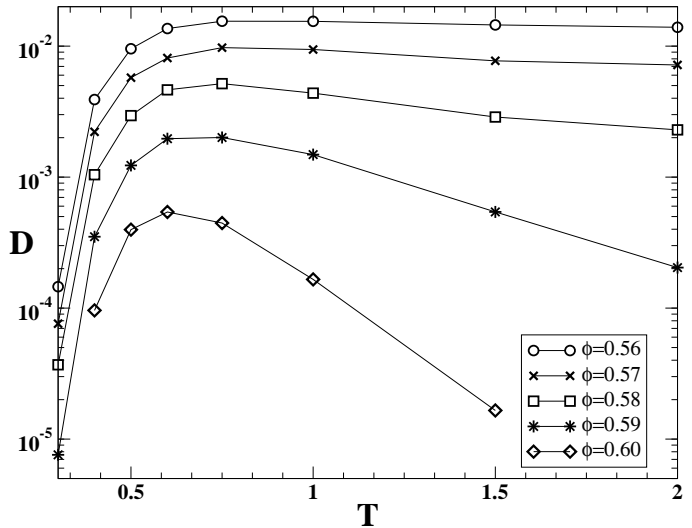}
           }
\vspace*{0.5cm}
\caption{
  As in Fig.\protect\ref{fig:diffusivity} but, in this case, the
  diffusion costant has not been normalized, and it is plotted against
  temperature along isochores between $\phi=0.56$ and $\phi=0.60$. The
  maximum in the bare diffusivity becomes more evident (almost $2$
  orders of magnitude) as one moves in the more reentrant part of the
  $(\phi,T)$-plane. }
%\vspace*{0.2cm}
\label{fig:diffmax}
\end{figure}

\begin{figure}
\centerline{\psfig{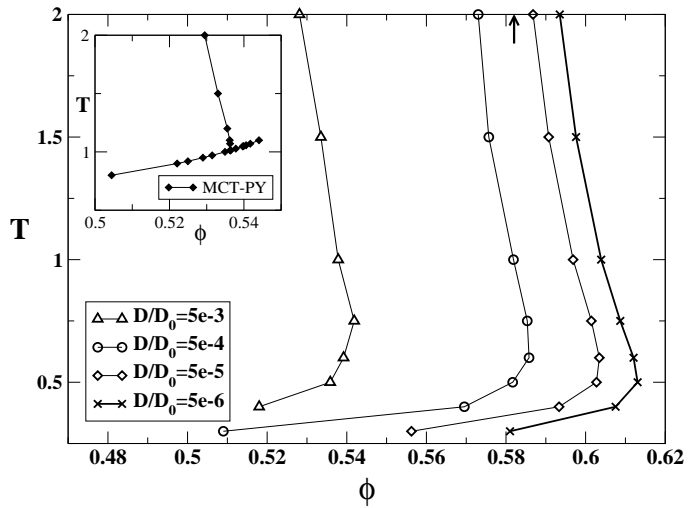}
           }
\vspace*{0.5cm}
\caption{
  Curves of iso-normalised-diffusivity $D/D_0$ in the $(\phi,T)$
  plane. We indicate with a vertical arrow the location for the hard
  sphere case of the packing fraction ($\phi \sim 0.582$) for the
  lowest iso-$D/D_0$ curve, i.e. $D/D_0=5e-6$. The inset shows the MCT
  prediction, calculated in \protect\cite{dawson00}, with as input the
  structure factor obtained using Percus-Yevick approximation. }
\label{fig:isodiff}
\end{figure}

We also note that, if we plot the bare diffusion coefficients, as
evaluated from the fit of the MSD, without normalising by $D^0$, the
reentrant behaviour is preserved. This can be an advantage, from an
experimental point of view, because it would allow to observe at the
same packing fraction for various temperatures the diffusivity firstly
increasing then decreasing again, without having to include the
thermal factor. We plot in Fig.~\ref{fig:diffmax} the not-normalised
diffusivities at fixed packing fractions, varying the temperature. The
appearance of a maximum, sharper with increasing density, is indeed
another manifestation of the reentrance. Extracting the values of
maximum diffusivity, it is possible to draw a `maximum diffusion' line
in the phase diagram. This could be interesting as, for example, in a
complex system like water, this line appears to play a very important
role in the understanding of the `metastable' part of the water phase
diagram \cite{water}.

To make contact with the theoretical results for the ideal glass
transition, we have extrapolated curves of normalised
iso-diffusivities, as for the mono-disperse case \cite{RComm02} and
represented them in Figure~\ref{fig:isodiff}. Of course, the limit
$D\rightarrow 0$ would correspond to the ideal glass line, as
calculated by MCT. We report the curves for normalised values varying
between $5\cdot 10^{-3}$ and $5\cdot 10^{-6}$, and in the inset we
present for comparison the ideal glass line as predicted by MCT for a
SW one-component system.

It is interesting to comment on the behaviour of the iso-diffusivity
curves. Depending on which diffusivity value is chosen, the most
reentrant point, i.e. the $(\phi,T)$-point characterized by largest
packing fraction with that diffusivity, changes. In the considered
range of diffusivities, its temperature varies from $0.75$, which also
corresponds to the most reentrant point for the mono-disperse case, to
$0.5$. However, data of Fig.~\ref{fig:diffusivity} allow to say that,
in the limit $D\rightarrow 0$, such point will be found at a finite
temperature between $0.4$ and $0.5$, since the reached packing
fractions are so large that it would not be possible, within the
trend, to go much beyond.

The reentrant behaviour, present in Mode Coupling calculations, is
then confirmed by simulation. It is clear that, since MCT
underestimates the effect of packing, we should not expect, a perfect
quantitative agreement between theory and numerical results. Indeed,
for example for the simple hard sphere case, the MCT critical glass
transition packing fraction is $\phi \simeq 0.516$ \cite{barrat}
whereas in experiments this has been shown to be $\phi \simeq 0.58$
\cite{vanmegen}. On the other hand, the same experiments have shown
that more accurate predictions can be expected for the behaviour of
dynamical quantities such as the density-density
correlation functions and the MSD.

\section{Results: Dynamics along Isotherms}

%{\bf HERE ARE SOME GENERIC COMMENTS --- \\
%1) IT WOULD BE NICE TO
%HAVE FIGURES FOR CORRELATORS AND MSD GROUPED IN THE SAME FIGURE.
%THIS WILL ALNMOST HALF THE NUMBER OF FIGURES.\\
%2) IT WOULD BE NICE TO HAVE IN EACH CORRELATOR FIGURE A LINE WITH
%THE ESTIMATED fq FROM THE FIT. \\
%3) I WOULD LOVE TO SEE THE COMPARISON WITH THE HS CASE (T=oo)
%FOR THE SAME BINARY MIXTURE. THIS COULD BE DONE WHILE
%WE WAIT FOR THE REPORT IF KEN WANTS TO RUSH.}\\

%%%%%%%%%%%%%%%%%%%%T=2.0%%%%%%%%%%%%%%%%%%%%%%%%%%%%%%%%%%%%%%%%%%%%
We start by examining the results for the isotherm $T=2.0$. Here
particles have sufficient thermal energy to escape the attractions and
the resulting dynamics appears to be quite similar to that expected
for ordinary hard spheres. However, the effect of the attraction,
thought not changing the general behaviour, is to enlarge the liquid
part of the phase diagram toward packing fractions already quite
larger than the typical (one-component) hard sphere value,
i.e. $0.58$. At this temperature, indeed, the system behaves as a
fluid at least up to a packing fraction $\phi=0.595$, where the time
limit of our simulation is reached. Close to this limit value, the
diffusivity decreases almost two orders of magnitude for a variation
of $1\%$ in packing fraction. 

\begin{figure}
\centerline{\psfig{figure=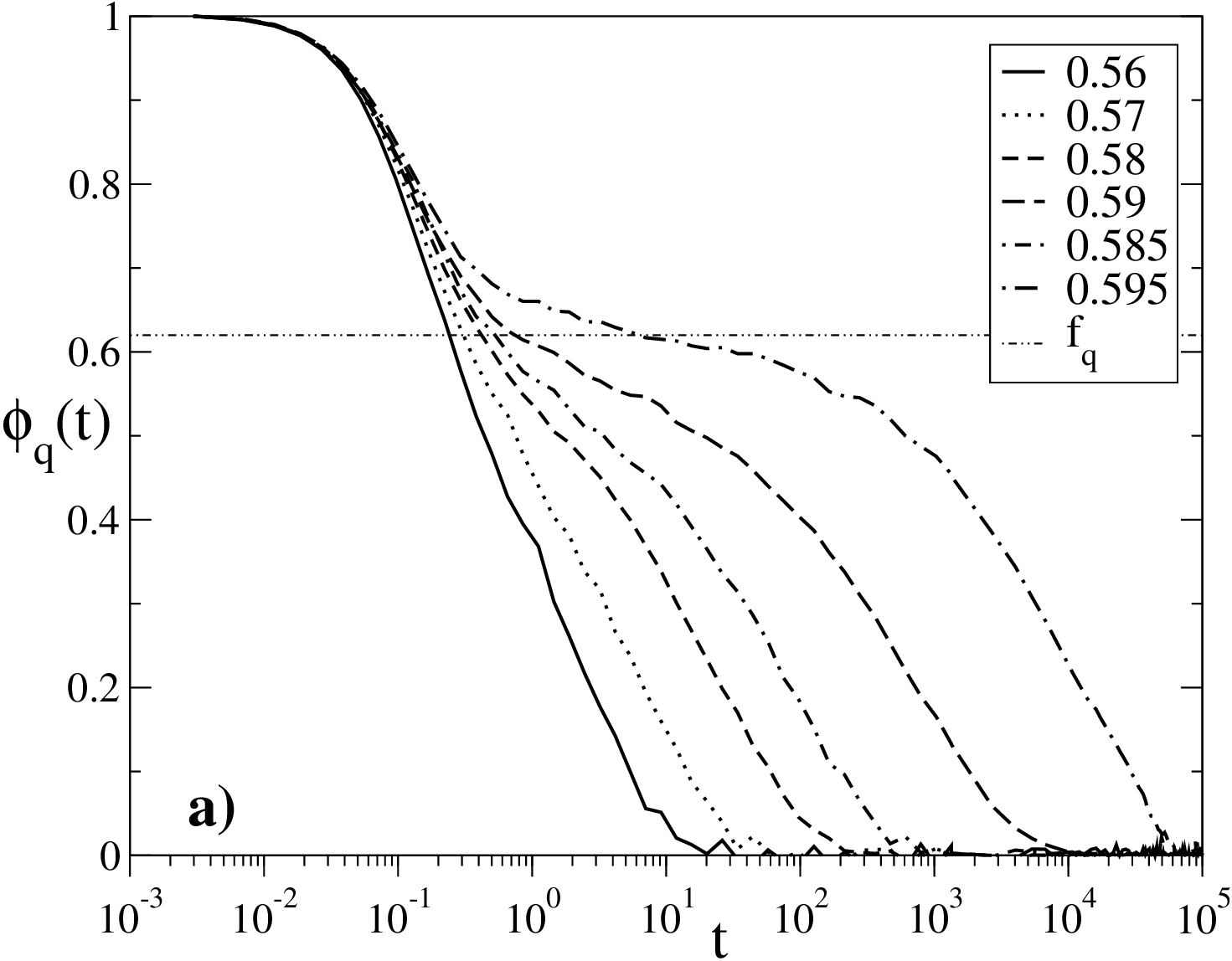,width=9.cm,clip=,angle=0.}
           }
  \vspace*{0.5cm}
\centerline{\psfig{figure=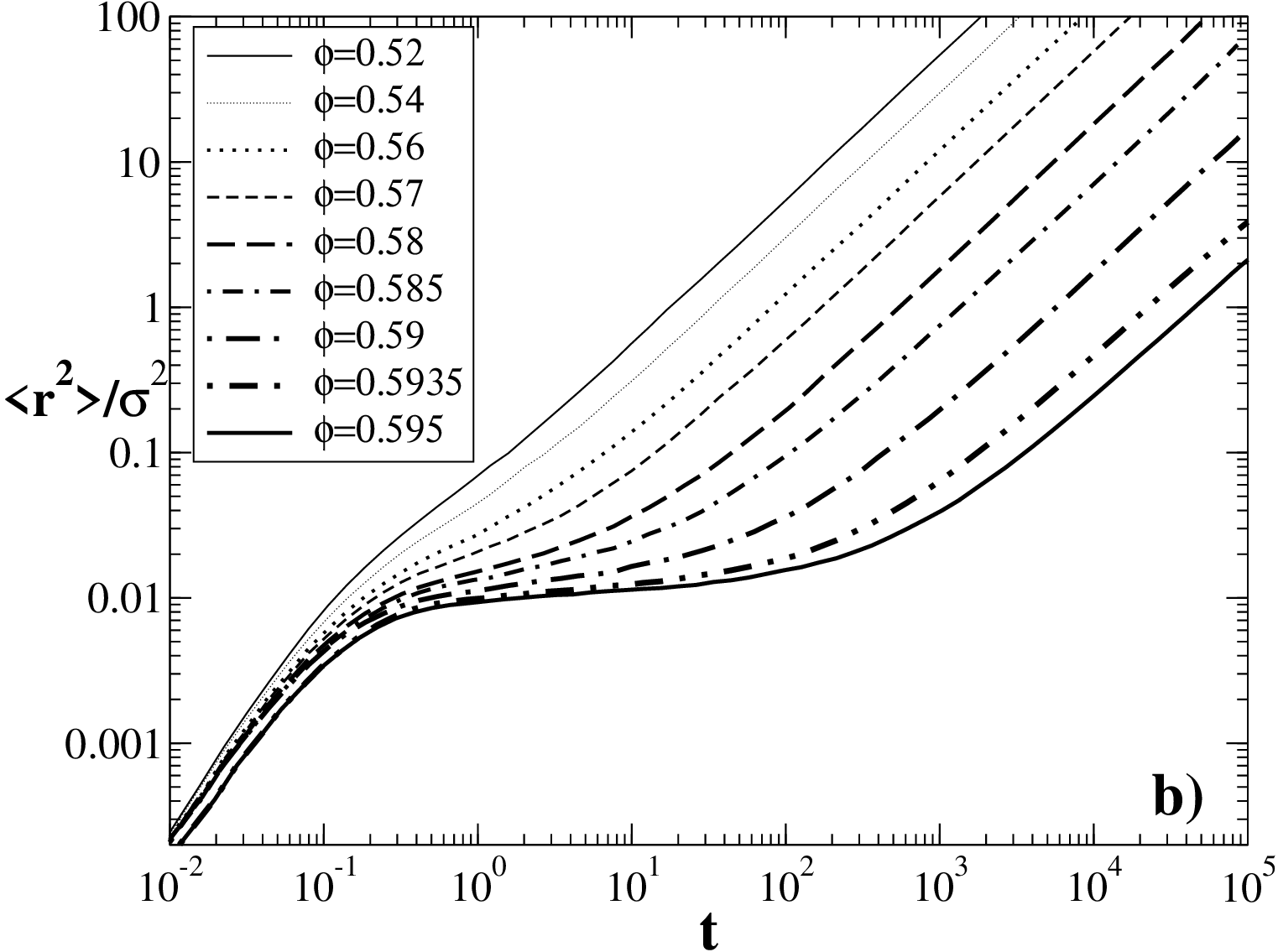,width=9.cm,clip=,angle=0.}
           }
\vspace*{0.5cm}
\caption{a)  
  Normalised density-density correlation function along the isotherm
  $T=2.0$ for different packing fractions.  The $q$-vector displayed
  in this picture is slightly larger then the one for the first peak
  of the static structure factor.  The horizontal line represents the
  correspondent non-ergodicity parameter,
  as extrapolated from the fit (see text);\\
  b) Mean square displacement along the isotherm $T=2.0$ for the same
  densities as in figures \protect\ref{fig:corre-2}(a).}
\label{fig:corre-2}
\end{figure}

In Figure~\ref{fig:corre-2}(a) we report the time-dependent
density-density correlators at increasing packing fraction, up to the
closest to the ideal glass transition. The dramatic decrease in
diffusivity is reflected in the behaviour of the correlators by the
formation of the typical two step relaxation process near the arrest,
described above.  Similar behaviour can be also observed in the
behaviour of the MSD displayed in Fig.~\ref{fig:corre-2}(b) in unity of
$\sigma^2$.  We note, however, that the height of the plateau of
approximately $10\%$ of particles diameter is consistent with
Lindemann's melting criterion \cite{lindemann}.

%In Fig.... we present the normalized diffusivity along the
%isotherm $T=2.0$ together with the curves that represent the a fit
%obtained using the functional form \ref{MCT-diffu}. It is evident that
%the data are well represented by such a power law. 

The diffusivity data can be fitted with the MCT power-law behaviour
(Eq.~\ref{MCT-diffu}). This holds sufficiently close to the ideal
glass transition, thus we considered relevant for the fit only those
points for which a clear $\alpha$-relaxation process was evident.
Doing so, we found $\gamma\simeq 3.6\pm0.4$. This value is already
much larger than the typical one-component hard sphere value predicted
by MCT, i.e. $\gamma^{HS}=2.58$ \cite{barrat}. In the case of our
particular binary mixture, in the hard sphere case, whose diffusivity
behaviour is also reported in Fig.~\ref{fig:diffusivity}, we found a
value of $\gamma=2.9\pm0.2$.  The uncertainty on the exponent is due
to the variation it gets when considering only the points closest to
the transition. We note that in reference
\cite{puertas02}, the case reported brings a value of $\gamma=3.03$.
Thus, we are in a situation clearly closer to a higher order
singularity, being the correspondent other MCT exponents for this
value, respectively $b=0.35$ and $\lambda=0.874\pm0.03$.  It is
perhaps important to stress at this point that the value of $\gamma$
obtained with this procedure can be slightly wrong due to the
difficulty to get close enough to the ideal glass transition with
numerical simulations \cite{kob-97}.  We will use the so-calculated
value of $b$ in the next paragraph to fit the behaviour of density
correlators along the iso-diffusivity curve $D/D_0=5\cdot10^{-6}$, to
give an idea of the behaviour of quantities of interest along the
ideal glass transition line.

%%%%%%%%%%%%%%%%%%%%%%%%%%%%%%%%%%T=1.5e1.0%%%%%%%%%%%%%%%%%%%%%%%%%%%%
The case $T=1.5$ shows a behaviour completely analogous to $T=2.0$. We
note however that the fit of the diffusivity with Eq.~\ref{MCT-diffu}
gives in this case the exponents $\gamma=3.8\pm0.5$, $b=0.325$ and
$\lambda=0.888$.  The increase in the value of $\lambda$ is expected
since we are getting closer to the reentrant region, and consequently
to the singularity.  The same trend is observed for $T=1.0$ with
$\gamma$ still increasing up to about $3.96\pm0.12$, and $\lambda$
reaching the value of $0.896$.

\begin{figure}
\centerline{\psfig{figure=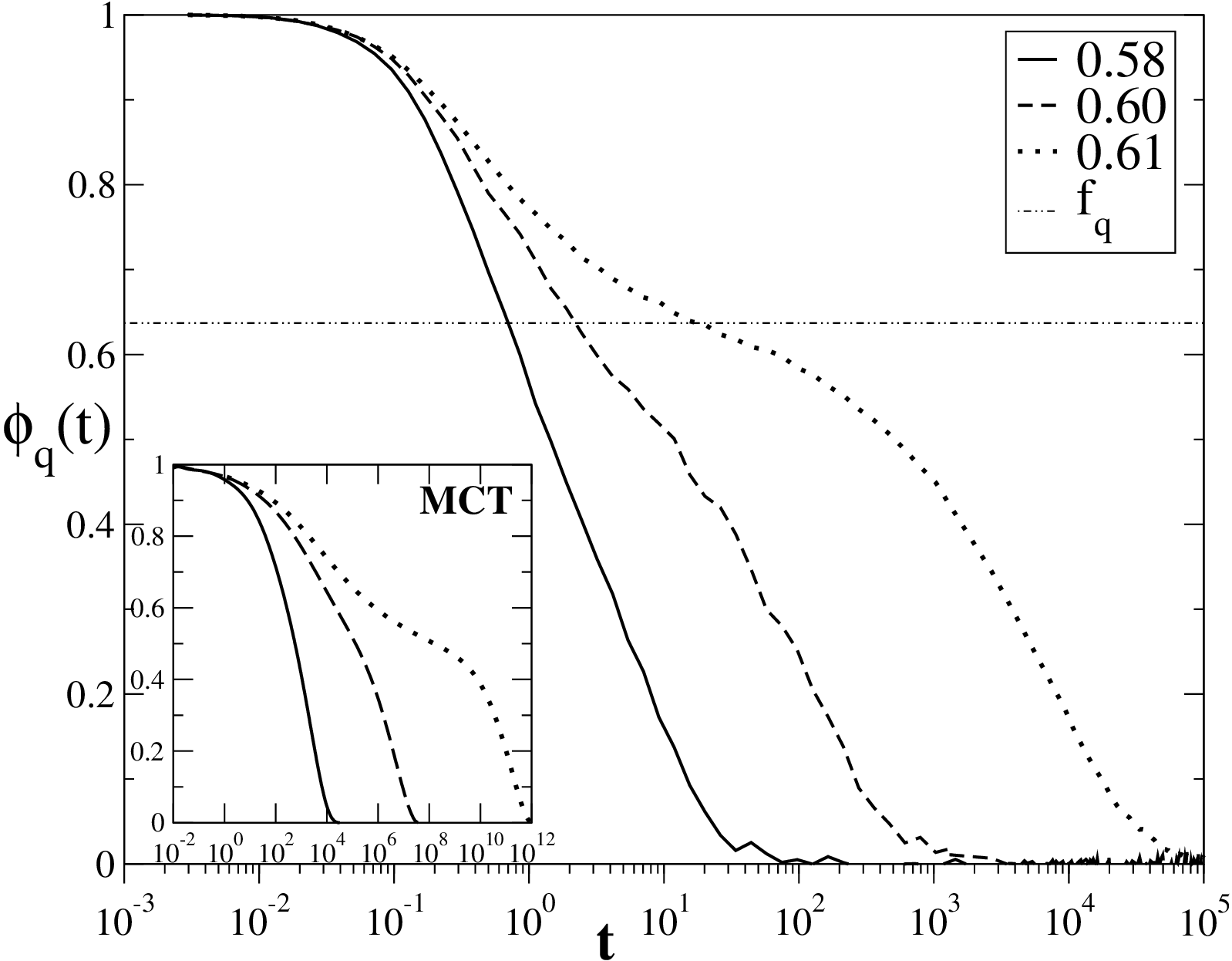,width=9.cm,clip=,angle=0.}
           }
\vspace*{0.5cm}
\caption{
  Same as in Fig.\protect\ref{fig:corre-2}(a) for $T=0.75$. The reported
  densities in this case have been chosen to evidence the analogy with
  the MCT predictions of Fig.11 in Ref.\protect\cite{dawson00},
  reproduced in the inset for comparison.}
\label{fig:corre-0.75}
\end{figure}

%%%%%%%%%%%%%%%%%%%%%%%%%%%%%%%T=0.75%%%%%%%%%%%%%%%%%%%%%%%%%%%%%%%%%%%
At $T=0.75$ a new interesting behaviour appears. Indeed, in the
density correlators a logarithmic decay starts to emerge. As shown in
Fig.~\ref{fig:corre-0.75}, some state points display a logarithmic
decay (see curve for $\phi=0.58$ in the figure) for almost the whole
relaxation process, i.e. after the microscopic relaxation up to
complete decay.  The shape of this logarithmic behaviour appears quite
different from the one found in reference \cite{puertas02}. On the
other hand, it reminds quite closely the shape of MCT correlators near
the $A_3$ singularity for the $3\%$ square well potential reported in
Figure 11 of Ref. \cite{dawson00}, which for comparison is reported in
the inset in Fig.~\ref{fig:corre-0.75}. Upon increasing density, and
so getting closer to the glass transition, the relaxation changes to
the usual two-step form, clearly indicating a similar situation of our
isothermal path to that indicated in the inset of Fig. 11 cited above.
Thus, the higher order singularity dominates the dynamics at smaller
packing fractions, but when one gets sufficiently close to the glass
transition, a conventional $A_2$ singularity is met, and this causes
the restoration of the typical $\alpha$-relaxation.  By considering
only those packing fractions, when at least the beginning of the
$\alpha$-relaxation can be observed (i.e. $\phi >0.6$), we can fit the
diffusivity again with Eq.~\ref{b-law}, obtaining the extremely high
value $\gamma \sim 5.1$, corresponding to the value of $b\sim 0.25$
and $\lambda=0.937$.  We cannot here estimate the error, due to the
small number of points available (only five for three parameters for
the fit), however, even if not so precise, it clearly indicates we are
approaching a higher order singularity.  As we shall see by looking at
the shape of the non-ergodicity factors in the next paragraph, we can
anticipate that we are meeting the $A_2$ line along its repulsive
branch, again as in the inset of Fig.~11 of ref.\cite{dawson00}, but
probably at a slightly higher temperature.

\begin{figure}
\centerline{\psfig{figure=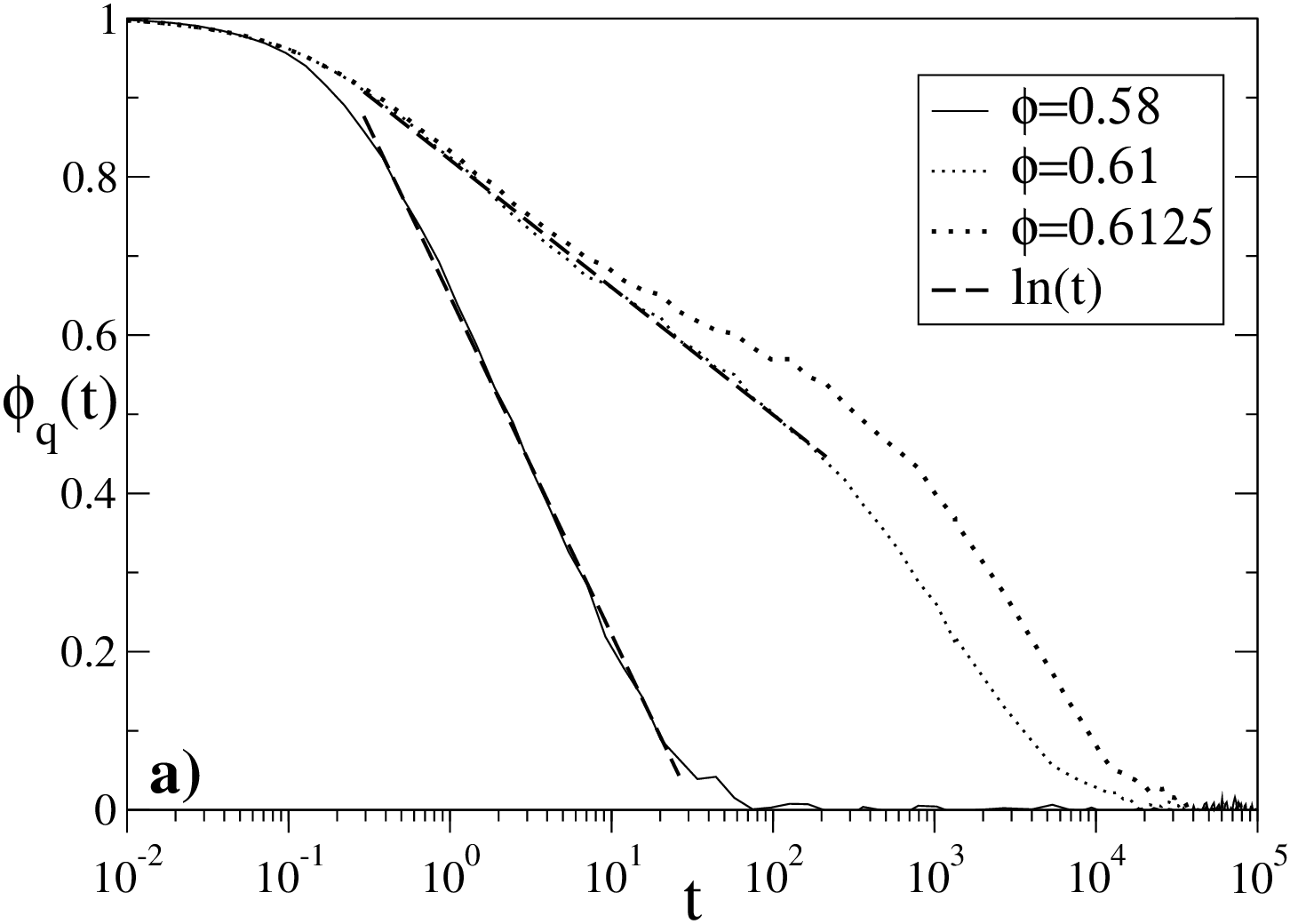,width=9.cm,clip=,angle=0.}
           }
\vspace*{0.5cm}
\centerline{\psfig{figure=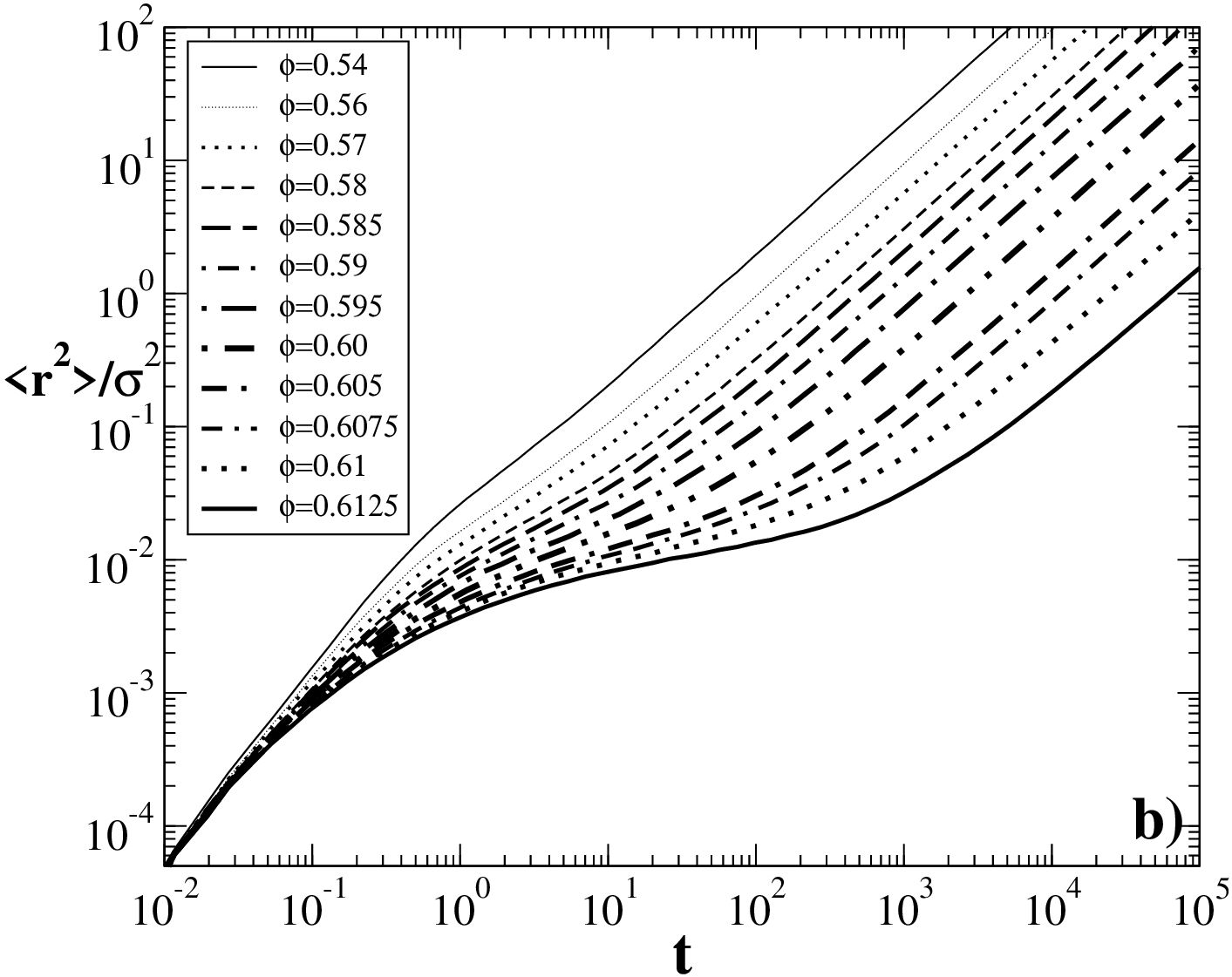,width=9.cm,clip=,angle=0.}
           }
\vspace*{0.5cm}
\caption{a) 
  Same as in Fig.\protect\ref{fig:corre-2}(a) for $T=0.6$. The dashed
  lines represent fits with logarithmic laws. They are displayed to
  show the presence of a logarithmic decay and the mechanism of its
  disappearance in the proximity of an $A_2$ singularity (see text for
  further details). They are reported as a guideline to the eye, and
  not to extrapolate any fit parameters.\,\,\, b) As in
  Fig.\protect\ref{fig:corre-2}(b) for $T=0.6$.}
\label{fig:corre-0.6}
\end{figure}

%%%%%%%%%%%%%%%%%%%%%%%%%%%%%%%%%%T=0.6%%%%%%%%%%%%%%%%%%%%%%%%%%%%%%%%%%%
The last observation derives from the analysis of the $T=0.6$
isotherm.  The correlators, reported in Figure~\ref{fig:corre-0.6}(a),
show an even closer behaviour to the one predicted by MCT in
\cite{dawson00} (again Fig.11, i.e. inset of
Fig.~\ref{fig:corre-0.75}). However, at this temperature, the two
competing singularities must be so close to each other that a clear
$\alpha$-relaxation does not take place within the reach of our
simulation, i.e. the logarithmic behaviour remains always very
important, and even at higher packing fractions it is clearly
observable before the $\alpha$-process takes over. The interesting
feature emerging is that, in all the cases considered the logarithmic
behaviour does never extend for much more than $3$ and a half decades
in time. This arises because, in the present topology of the phase
diagram, one is either to close to the $A_2$-singularities to observe
a pure logarithm, or is too far from the glass transition and thus the
relaxation time is generally not too large. Indeed, this behaviour is
strongly supported again by the theoretical calculations in
\cite{dawson00}.

It is to note that at this temperature we are not able to convincingly
fit the power-law density dependence of the diffusivity.  Indeed, if
one forces the fit on the points, one find exponents strongly
dependent on the selected $\phi$ range.  A possible explanation for
this data sensitivity to $\phi$ can be found in the competition
between type $A_2$ and $A_3$ or $A_4$ dynamics. In such condition,
only a comparison with a full MCT solution (as opposed to an
asymptotic prediction) may help in rationalizing the density
dependence of diffusivity.  In agreement with the previous
observations, also the MSD, represented in Figure~\ref{fig:corre-0.6}(b)
starts to show deviations from usual type $B$ behavior.  Indeed, a
clear flat region does not appear, though dynamics are significantly
slow.  What can be observed is a slight deviation from the flat region
found at higher temperatures, which will become more and more evident
at lower temperatures.  No clear localization length can be found.
Attraction at this temperature has become quite relevant. It is again
a sign of very strong competition between different singularities,
between attractive and repulsive cages.

\begin{figure}
\centerline{\psfig{figure=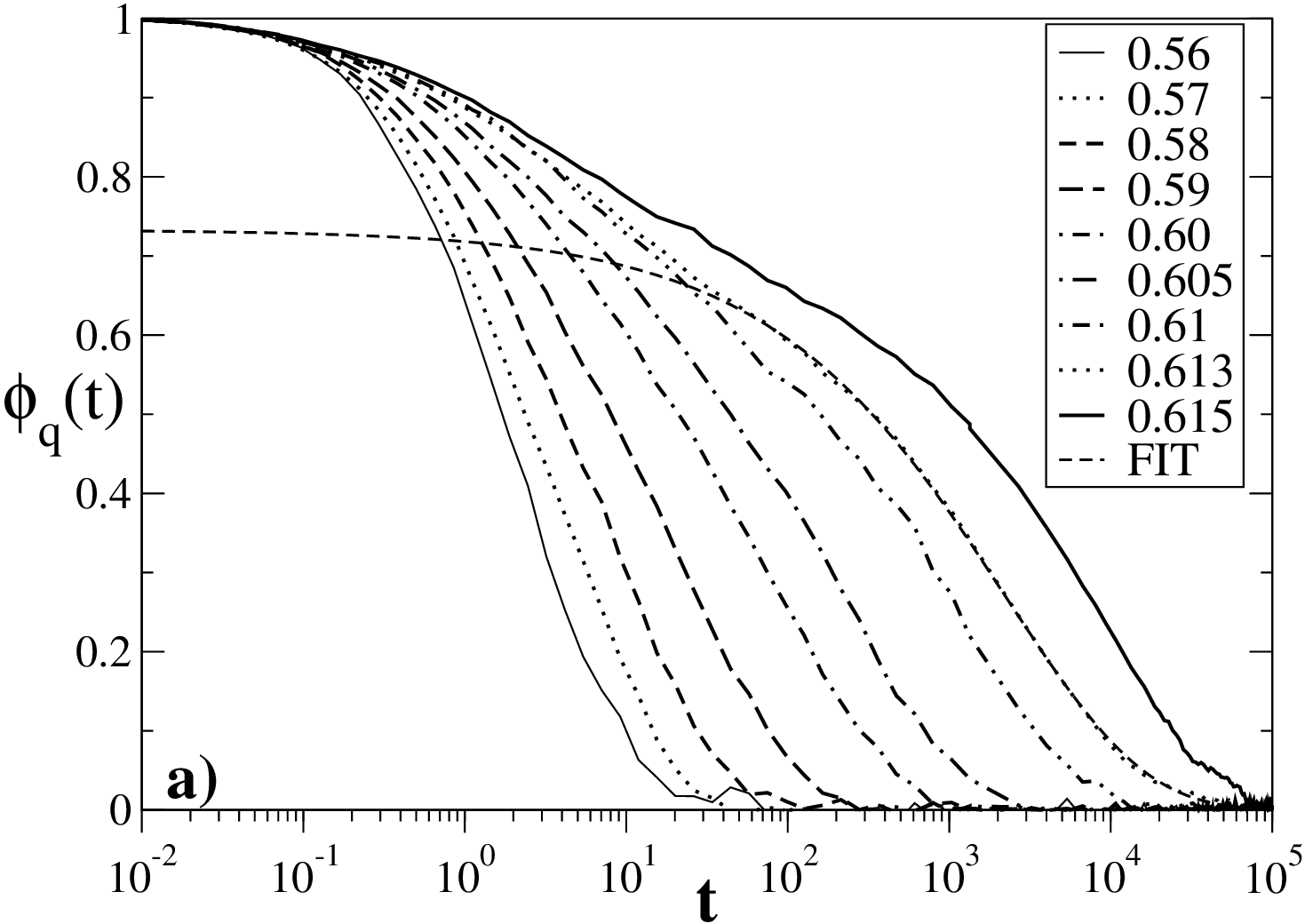,width=9.cm,clip=,angle=0.}
           }
\vspace*{0.5cm}
\centerline{\psfig{figure=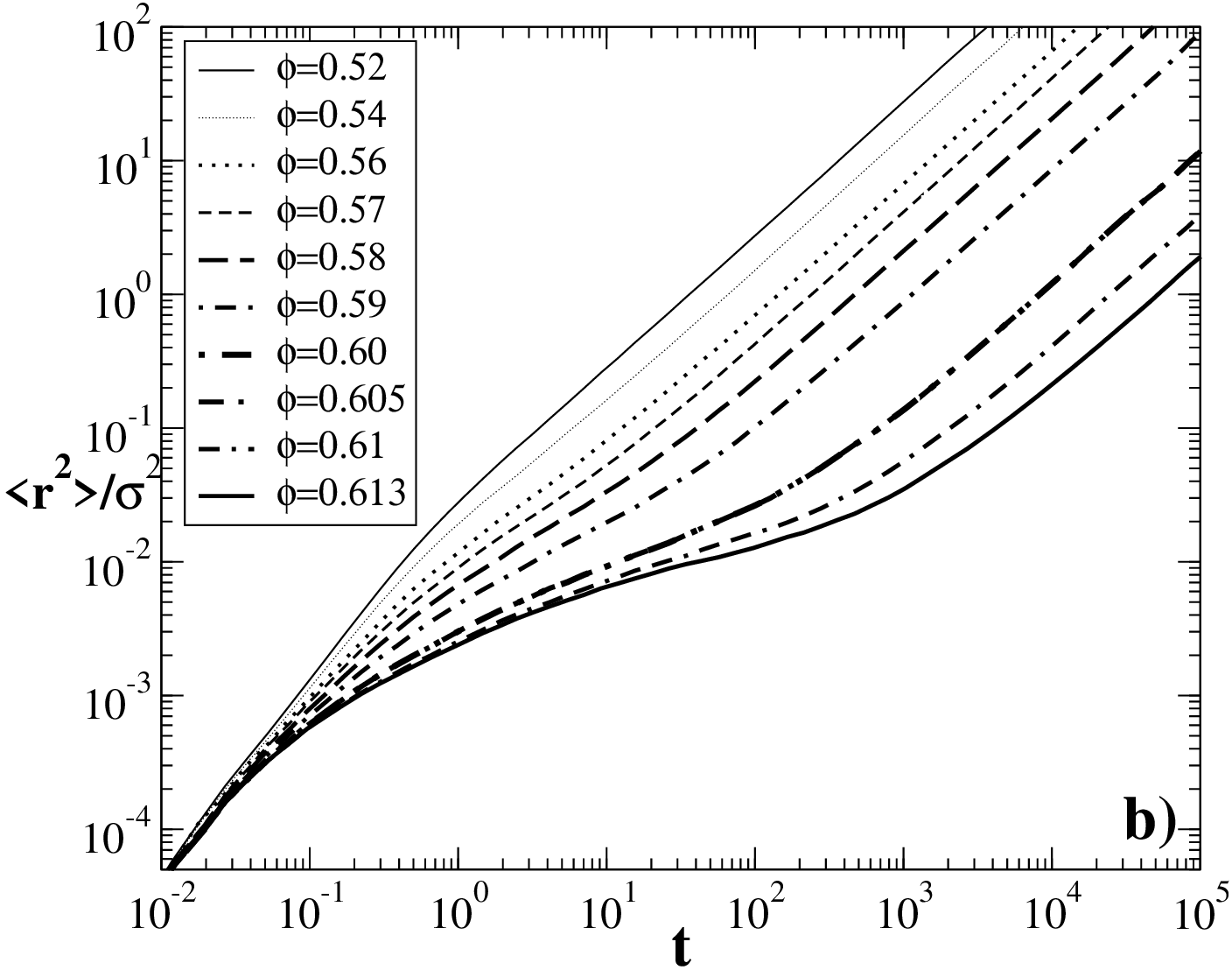,width=9.cm,clip=,angle=0.}
           }
\vspace*{0.5cm}
\caption{a) As in Fig.\protect\ref{fig:corre-2}(a) for  $T=0.5$. 
  The dashed line is a fit of one of the correlators with a stretched
  exponential, which we used to extrapolate the non-ergodicity
  parameter $f_q$, and it is shown to display the goodness of the
  fit.\,\,\, Parameters of the stretched exponential fits are reported
  in Fig.\protect\ref{fig:beta-tau}. b) As in
  Fig.\protect\ref{fig:corre-2}(b) for $T=0.5$.}
\label{fig:corre-0.5}
\end{figure}

%%%%%%%%%%%%%%%%%%%%%%%%%%%%%%%%%%T=0.5%%%%%%%%%%%%%%%%%%%%%%%%%%%%%%%%%%%
Upon further decreasing temperature, we enter the most delicate region
of the phase diagram. Indeed at $T=0.5$, as for $T=0.6$, it is not
possible to find any MCT exponents, and the situation gets even worse
in interpreting the behaviour of the correlators. These are plotted in
Figure~\ref{fig:corre-0.5}(a). Indeed, no clear behaviour, at all
densities, either logarithmic or of type $\alpha$, can be
individuated, and no evident plateau value ever arises.  We note that
the long-time decay of density correlators can be represented by a
stretched exponential, but with very low exponents $\beta_q$, as shown
in Figure \ref{fig:beta-tau}.  In the MSD, reported in
Figure~\ref{fig:corre-0.5}(b), the phenomenon present at the previous
temperature, becomes more accentuated.  Even the slowest studied state
point is far from being asymptotic, and the MSD presents a clear
transient region.

%%%%%%%%%%%%%%%%%%%%%%%%%%%%%%%%%%T=0.4%%%%%%%%%%%%%%%%%%%%%%%%%%%%%%%%%%%
The case where the anomalous dynamics and the interplay between
different singularities is fully displayed is offered by the $T=0.4$
isotherm. The correlation functions, shown in Figure
\ref{fig:corre-0.4}(a), are rather peculiar. Even the long time limit is
far from being rationalized in term of stretched exponential decay.
The MSD behaviour, shown in Figure~\ref{fig:corre-0.4}(b) is also quite
intriguing.  The MSD transient behaviour is now evidently of a
peculiar type.  Indeed, for about $4$ decades in time, it shows a
dependence which can be quite accurately described by a power-law
behaviour, i.e.
\begin{equation}
\langle r^2\rangle \sim t^{x}
\end{equation} 
We estimated via a fit $x \simeq 0.44$.  A similar behaviour has been
found in the MCT study of polymeric systems \cite{chong02}, for
displacements varying from the typical localization length of
hard-sphere-like cages to end-to-end distance. The analogy with the
polymeric systems, where permanent bonds are present (in a sense close
to the attractive cages at this very low temperature), can be a guide
to a deeper understanding of this regime.

The strong effects that we find at this value of temperature seem to
suggest that along this isotherm the system approaches the closest
point to the singularity, even if we do not know yet on which side
(attractive or repulsive) of the glass line it will be located.  To
understand the nature of the dynamics which takes place here further
investigations are needed, and maybe a more complete analysis of the
correlators, and a comparison with full solutions of the MCT
equations.

%%%%%%%%%%%%%%%%%%%%%%%%%%%%%%%%%%T=0.3%%%%%%%%%%%%%%%%%%%%%%%%%%%%%%%%%%%
Finally we analyze the last isotherm, corresponding at $T=0.3$. This
being a very low value for the system to equilibrate, data are not so
clean as for the other cases, also because here one needs to study
slower points with respect to the other temperatures in terms of bare
diffusivities, to reach the same values of normalised ones.  However,
despite these technical difficulties, we find more transparent results
in terms of conventional MCT interpretations, i.e. we can identify the
development of a two-step process typical of an $A_2$ singularity,
both for the correlators and for the MSD than in the previous case.

Indeed, observing the correlators in Figure~\ref{fig:corre-0.3}(a), it
is clear that, close enough to the transition, they present the
development of a plateau, and thus, an $\alpha$-relaxation process, as
shown in the inset of the figure for various $q$-vectors.  However,
this plateau is extremely high, and clearly indicates that we are now
undoubtedly in the condition of attractive glasses. It is of deep
interest to note the analogy of this behaviour with the one that has
been found in the study of `strong' gels \cite{emanueladelgado}, even
if this should be inspected at lower densities also.  Despite this
clear behaviour, even at this temperature, it is not possible to
evaluate unambiguous power-law exponents from the diffusivity.  Figure
\ref{fig:corre-0.3}(b), represents the evolution of MSD at this
temperature. Here, a signature of a localization length, smaller than
the one in the high temperature cases, starts to develop. Indeed, an
indication of a plateau is observable around {\bf $\langle
  r^2\rangle=0.0006 \cdot \sigma^2$}. The corresponding localization
length is of the order of $\Delta$, supporting the interpretation that
at this temperature the relevant localization length has become the
attractive well.  In this respect, one can interpret the sub-diffusive
behavior of the MSD discussed at $T=0.4$ as a cross-over effect
between the hard-sphere and attractive well different localization
lengths, in a similar fashion to what has been obtained for polymeric
systems \cite{chong02}.

\section{Results: along the iso-$D/D_o$ curve}

We now focus on studying the behaviour of correlators and MSD, and
other quantities along the iso-(normalised)-diffusivity, i.e.
iso-$D/D_o$, curve $D/D_0=5\cdot10^{-6}$, shown in
Fig.~\ref{fig:isodiff}, which represents our closest available
representation of the ideal glass transition line. The aim of this
study is to give clear evidence of the existence of two distinct
glassy states, attractive and repulsive. Also, it aims to connect even
more closely this simulation to the MCT calculations, which also were
performed in a similar fashion, along the ideal glass lines, in
ref.\cite{zaccarelli01}.

We start by representing the behaviour of density correlators along
the iso-$D/D_0$ line in Figure~\ref{fig:corre-linea}.  The curves
here represented, having equal diffusivity, also have the same
normalized relaxation time. Thus, we can clearly see the change in the
decay which takes place, upon decreasing temperature, from a marked
$\alpha$-relaxation at higher temperature to the extremely slow decay
of $T=0.4$, passing through the intermediate regimes between $T=0.75$
and $T=0.5$. Of course, here, no evident logarithmic behaviour can be
observed, due to the proximity to the $A_2$ transition, as discussed in
the previous paragraph.

Next, we report the MSD behaviour along the line in Figure
\ref{fig:msd-linea}.  Despite the larger statistical error at $T=0.3$,
we display them as an important part of the whole picture.
Thus, here, we can clearly observe the change in the diffusion
process. The first evident thing to note is the big gap, of almost 2
orders of magnitude, in the localization length of particles, which,
as discussed above, characterises the size of cages around particles.
Clearly this fact can be used as a justification to speak of
`attractive' cages, opposed to normal cages, intended as simple
occupation of the available space.  In the attractive cages, the
average distance between particles is much smaller, the lower the
temperature. This is the first clear signal of different structure in
the glass formation.

Also, we can also examine more carefully the modification of the
plateau present at higher temperatures, in the transient regime.
Indeed, increasing the attraction, this tends to bend downwards until
a sort of `saturation' between the two competing mechanisms
(attraction and packing) take place, corresponding to the
sub-diffusive behaviour of $T=0.4$. After this point, attractions
become dominant, and the curve starts to bend upwards. This might
suggest that, going at even lower diffusivities, the MSD would display
a similar plateau as for high temperatures at roughly $\langle
r^2\rangle \sim 0.0007\cdot \sigma^2$ \cite{nota0.3}, which means roughly a
localization length of $2.6\%$ of the particle diameter, i.e.
comparable with the width of the attractive well of the model,
confirming our conjectures on the formation of attractive cages, or,
to use a another expression, bonds. However, to gain further evidence
on how these mechanisms really happen and evolve in the system, a
specific study of configurations in terms of average distance, sizes
of clusters, and heterogeneities in general should be performed, and
this is beyond the aim of the present work. We note that a similar
figure, showing the behaviour of MSD with attraction,has appeared in
\cite{science02}, but not all of these considerations could be made
there, due to the distance from the transition.

\begin{figure}
\centerline{\psfig{figure=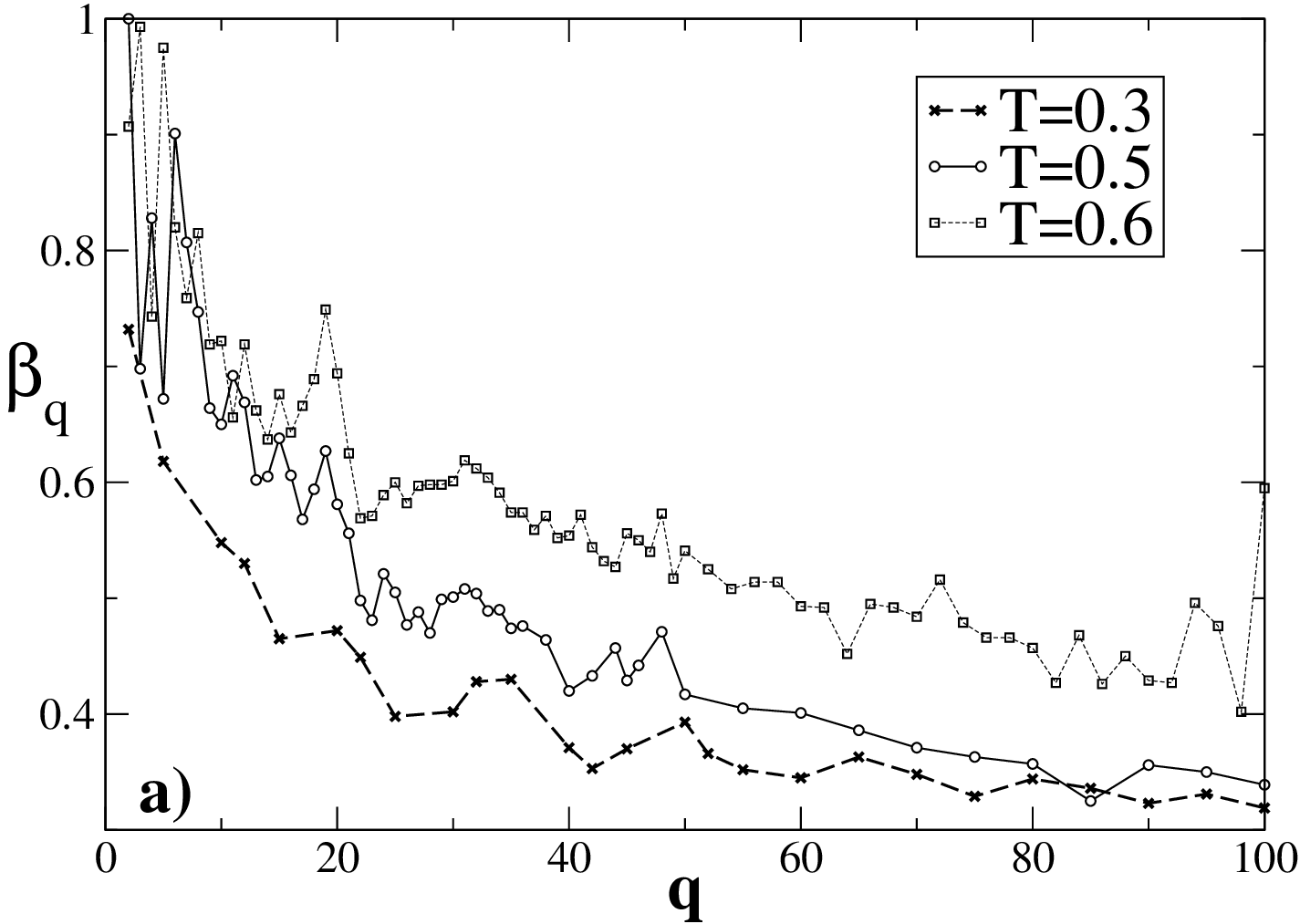,width=9.cm,clip=,angle=0.}
           }
\vspace*{0.5cm}
\centerline{\psfig{figure=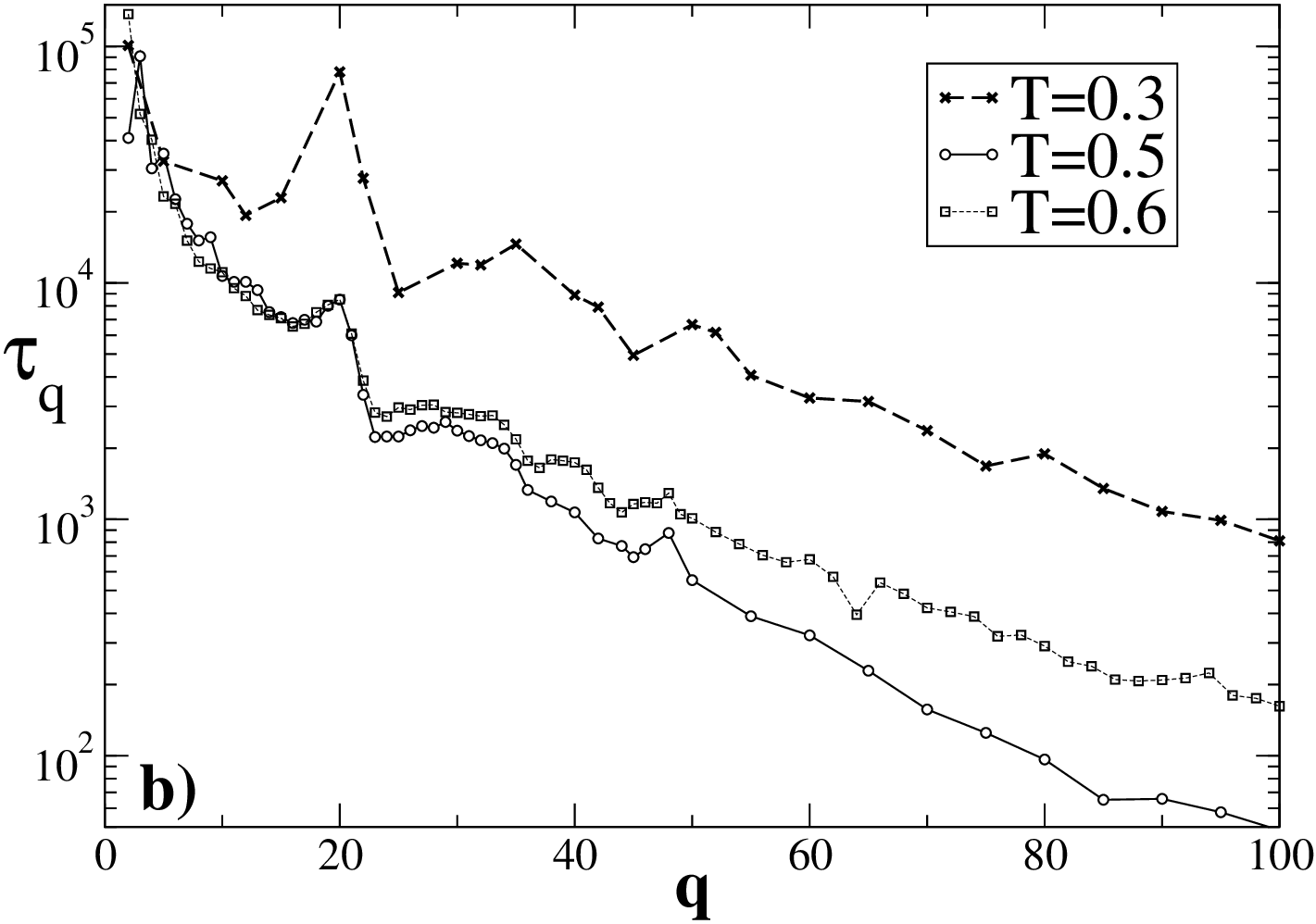,width=9.cm,clip=,angle=0.}
           }
\vspace*{0.5cm}
\caption{a)
  Exponent $\beta_q$ as a function of $q$ obtained from the fit of the
  density-density correlation function with the stretched exponential
  in Eq.\protect\ref{stretched} for temperatures $T=0.6,0.5,0.3$. The
  values found are always very small, indicating a very slow
  relaxation.\,\,\, b) As in a) for the relaxation time parameter
  $\tau_q$ of Eq.\protect\ref{stretched}.  Interestingly, a peak
  corresponding to the $q-$value of the static structure factor
  develops, as one lowers the temperature.}
\label{fig:beta-tau}
\end{figure}

\begin{figure}
\centerline{\psfig{figure=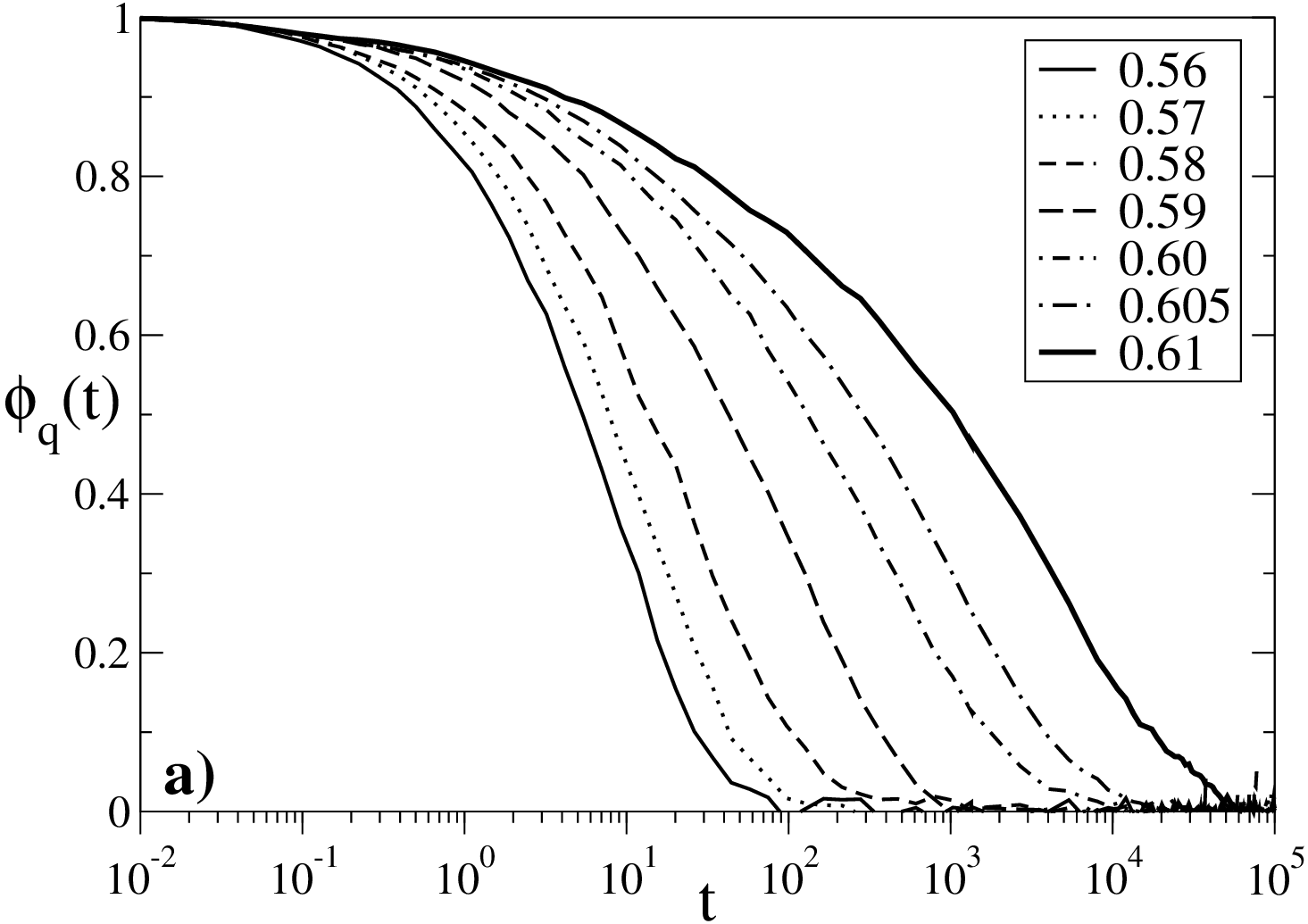,width=9.cm,clip=,angle=0.}
           }
\vspace*{0.5cm}
\centerline{\psfig{figure=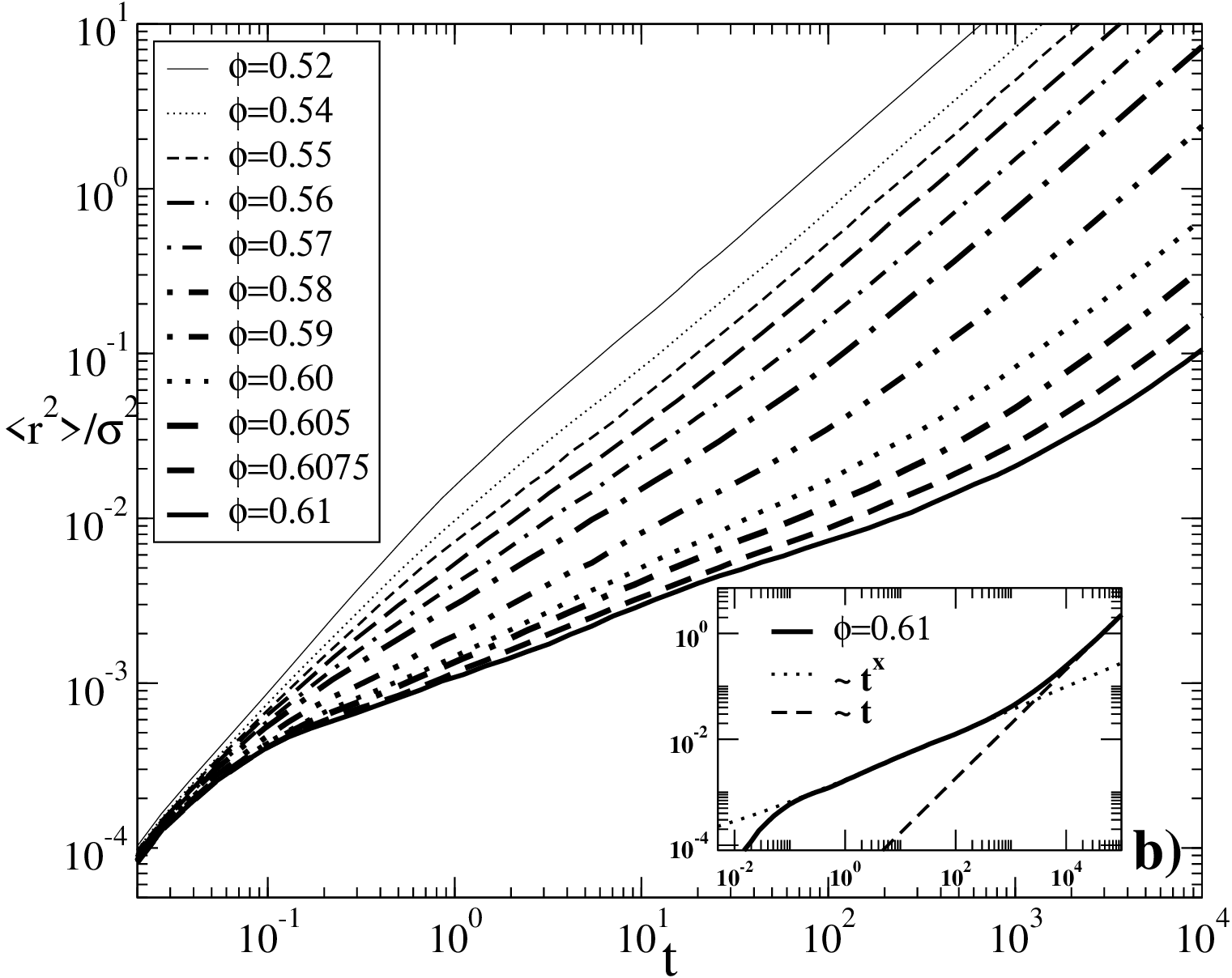,width=9.cm,clip=,angle=0.}
           }
\vspace*{0.5cm}
\caption{a) As in Fig.\protect\ref{fig:corre-2}(a) for  $T=0.4$. \,\,\, 
  b) As in Fig.\protect\ref{fig:corre-2}(b) for $T=0.4$. In the inset
  the fit of the sub-diffusive and subdiffusive regime with a power
  law is shown (see text for details).}
\label{fig:corre-0.4}
\end{figure}

We now turn to evaluate the non-ergodicity factor $f_q$ along the
iso-$D/D_0$ curve. To do this, we have fitted the density correlators
at various $q$-vectors, and extracted the relevant parameters. Where
possible, i.e. where the power-law behaviour for the diffusivity in
(\ref{MCT-diffu}) was found to be valid, we used the power-law
described in terms of the $b$ exponent for the $\alpha$-relaxation of
equation (\ref{b-law}).  Thus, for $T=2.0$, we implemented the fit
with $b=0.35$, while respectively for $T=1.5,1.0,0.75$ we used
$b=0.32,0.31,0.25$. These values have been found very good for the
fits, always finding a $\chi^2$ of the order of $10^{-4}$ or less. For
lower temperatures, it was not possible to use this strategy and,
consequently, we had to use the approach of the stretched exponential
in Eq~\ref{stretched}, and use its amplitude as an estimate for
$f_q$. The parameters of the fits, i.e. the exponent of the stretching
$\beta_q$ and the relaxation time $\tau_q$, for the considered
temperatures $T=0.6,0.5,0.3$ are reported in Fig.~\ref{fig:beta-tau}
to display their $q-$behaviour. Even though the stretched exponential
law is not analytically justified, it is quite established in the
literature to use it as a fitting law for extracting the
non-ergodicity parameter \cite{kob-97}. In the case of $T=0.4$ also
this strategy did not work, as already discussed \cite{nota0.4}.

In Figure~\ref{fig:fq-linea} we show the so-calculated
$f_q$. Amazingly, from $T=2.0$ down to $T=0.6$, they all collapse onto
the same curves, giving a strong evidence of MCT predictions for the
repulsive glass \cite{zaccarelli01}.  Thus, the repulsive glass is
independent of temperature, and also this shows how the passage to
attractive glass intervenes quite sharply. For lower temperatures, the
glass becomes then attractive, and the non-ergodicity parameter starts
to be modified with temperature, becoming finite also at much larger
$q$-vectors \cite{zaccarelli01}.  
Despite some errors generated by the
stretched exponential fits at these low temperatures or the
data noise at $T=0.3$ a significant change in the shape and
width of $f_q$ is seen between $T=0.5$ and $T=0.3$.
It could be that the case at $T=0.5$ is quite sensitive to
the singularity, and thus it is a somehow intermediate case. 
A more detailed study, either theoretically or by considering
intermediate or even lower temperatures for smaller packing fractions,
 will be helpful for clarifying this issue. On the other hand, the
establishment of the existence of the two glasses along the line
appears to be definite by these results.
To  support this statement, we have plotted in Fig.~\ref{fig:fq-sq} both
the (partial) static structure factor $S_{AA}(q)$ (rescaled by a
factor of $2$ for having a better visualization of the figure) and the
non-ergodicity factor, respectively at the highest, $T=2.0$, and at
the lowest, $T=0.3$, temperatures studied in this work, so to compare
the most extreme cases of repulsive and attractive glasses. Firstly,
we note how the oscillations of the non-ergodicity factors follow the
ones in the structure factor quite closely in both cases (for the
$T=0.3$ case, the scale of the figure almost flattens these
oscillations, which are more evident in
Fig.~\ref{fig:fq-linea}). Also, the $S(q)$  presents the  typical features
we expected from theoretical calculations within MCT and integral
equations. Indeed, the repulsive case shows a marked first peak, which
is the main responsible for the glass transition, while the attractive
one possesses larger secondary oscillations, that constitute a signal
of the smaller cages already described above.  Thus, this contributes
to establish not only the existence of the two glasses, but also, and
most importantly, the two distinct mechanisms which drive the
glassification in the two cases, i.e. simply packing and localised
attraction.

\section{Conclusions}
In this paper, using molecular dynamics, we have studied the dynamical
arrest phenomena of spherical particles interacting via a square well
potential. The square well potential has been studied as one of the
simplest canonical models of solids liquids and gases for many years
\cite{alder,noro,elliot,vega,chang,rascon}.  Here we have extended the
models applicability to the domain of dynamical arrest, and glassy
phenomena. Previous predictions for dynamical arrest from mode
coupling theory for the square well potential are available
\cite{dawson00,zaccarelli01}, so direct comparisons are feasible.
 
By using a well-adjusted binary mixture, we have been able to extent
our previous preliminary investigations \cite{RComm02} much closer to
the arrest transition, accessing diffusion constants that are $4$
orders of magnitudes smaller than in the previous calculations.
Nevertheless, results on the lowest valued iso-diffusivity curve
available for the single-component system are very close to those for
the binary mixture, so we may consider the role of the second
component to be mainly the prevention of crystallization.
 
In that regime where repulsions dominate, we recover an 
ideal-glass-transition with power law scaling of the diffusion 
constant. We also find an attractive branch to the dynamical 
arrest where theory has predicted the presence of an 'attractive' 
glass. Where attractions and repulsions compete in nearly equal 
terms we find re-entrance in the arrest curves when plotted in 
units of the microscopic temperature dependence of the diffusion 
constant. However, fixed-density diffusion constants, plotted 
without any normalization, exhibit a maximum in the diffusion 
constant as temperature is increased. This maximum locates the 
re-entrant liquid where mobility is anomalously high. We consider 
that the essential features of re-entrance in this regime for the 
square-well potential now to be confirmed, in agreement with the 
predictions of theory \cite{dawson00}. 
 
We have studied also the evolution of the density-density correlation
functions (dynamical structure factors) and, independently, the
mean-squared distance traveled by particles in the vicinity of the
reentrant regime. As expected, where repulsive interactions dominate,
we find the classical arrest scenario in which plateaus develop in
both functions as arrest is approached \cite{goetze99}. These plateaus
indicate the development of an observable characteristic cage time,
and are quite typical of prediction by MCT for hard sphere
systems. When attraction begin to compete on equal terms, in the
vicinity of the re-entrant regime, the theory has predicted the
existence of an $A_3$ singularity embedded in the arrested region. It
is therefore difficult to access this singularity directly by
molecular dynamics, but the theory has indicated that there are
distinctive signatures of this singularity in the re-entrant fluid
phase, on approach to arrest. In particular, density correlators from
a suitable fixed-temperature cut of the phase diagram have an
interesting pattern of behavior in which the logarithmic behaviour
\cite{sperl02}, due to the embedded $A_3$ point, first begins to
dominate, and then crosses over to the conventional $A_2$ singular
behavior more commonly observed for normal MCT arrest. The density
correlators in the re-entrant regime clearly exhibit this phenomenon,
the pattern of evolutions being essentially in agreement with the
predictions of theory.
 
We may pause here to comment that we do not regard this complex 
cross-over behavior as a complication, but in fact as a rather 
delicate, and unusual signature of the whole re-entrant 
phenomenon, and interplay between $A_2$ and $A_3$ singularities. That 
the simulations would reproduce this is strong support for the 
detailed picture offered by theory. The behaviour of the 
mean-squared displacements are also quite unusual, and there is as 
yet no theoretical prediction for them in this regime. 
 
Finally, we are able to extract the non-ergodicity factors along the
arrest curve, for the lowest iso-(normalised)-diffusivity constant
curve. Again, in line with theoretical predictions we find strong
evidence of a transition from a repulsive glass to attractive glass
behavior, as indicated by the change in characteristic shapes of the
non-ergodicity factors. This is the first direct evidence of the
repulsive glass-to-attractive glass transition that has been predicted
by the theory, representing one of the most remarkable phenomena
associated with the system.
 
The theory suffers from strong systematic shifts of all the arrest 
curves in relation to the simulated ones, a phenomenon long known 
from the example of the hard sphere. However, in detail the 
theoretical predictions of re-entrant regime, with associated 
crossover to logarithmic singularity, and glass-to-glass 
transition has been confirmed by detailed molecular dynamics 
calculations. 
 
From the experimental point of view, there is accumulating evidence
that all the phenomena described here are robust, being relatively
independent of the details of the experimental system used to study
them \cite{bartsch,mallamace,science02}. The same is true of the
theoretical studies
\cite{bergenholtz99,fabbian99,dawson00,zaccarelli01,foffi02,merida} and
simulations \cite{puertas02,RComm02}. The square well potential is one
of the simplest examples one can study, and it is reassuring that it
exhibits the phenomena.
 
Our original prediction that in the dense regime, colloidal 
particulates system with short-ranged potentials could be 
described using ideas from dynamical arrest and glass theory now 
seems to  be strongly supported. Dense particle gels are 
thereby identified as an example of a new type of glass, or 
dynamically arrested phase. The implications are broader than the 
simple example studied, for it indicates that it may be possible 
to interpret many formerly disparate phenomena such as 
coagulation, precipitation, aggregation, and  gellation within the 
paradigm of dynamical arrest or glass theory. This is a 
fundamental soft of perception in the field of dense 
soft-condensed matter which may prove to be very fruitful in 
coming years.

This research is supported by the INFM-HOP-1999, MURST-PRIN-2000 and
COST P1. S.B. thanks the University of Rome and NSF, Chemistry
Division (Grant No. CHE0096892) for support.

\begin{figure}
\centerline{\psfig{figure=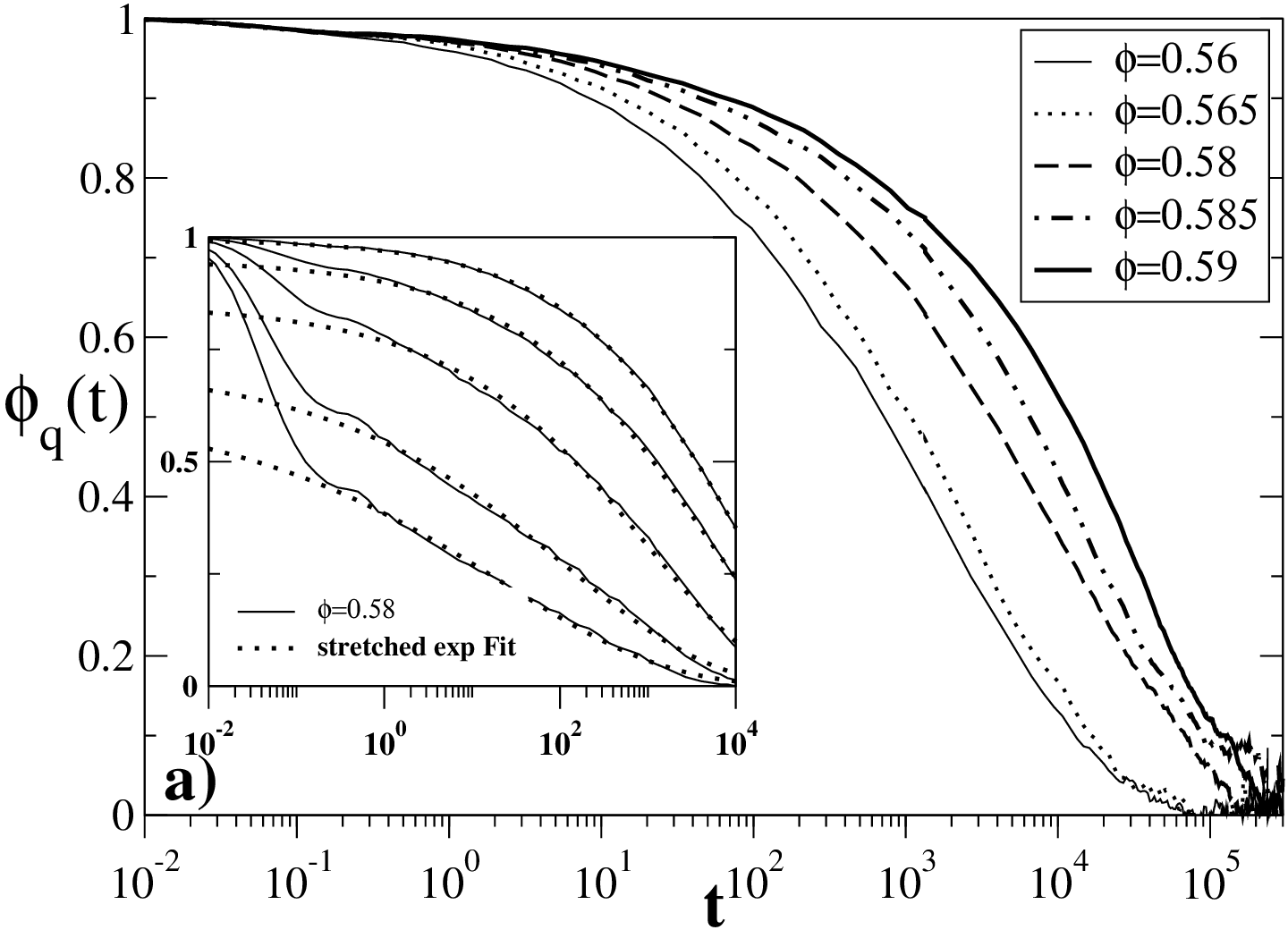,width=9.cm,clip=,angle=0.}
           }
\vspace*{0.5cm}
\centerline{\psfig{figure=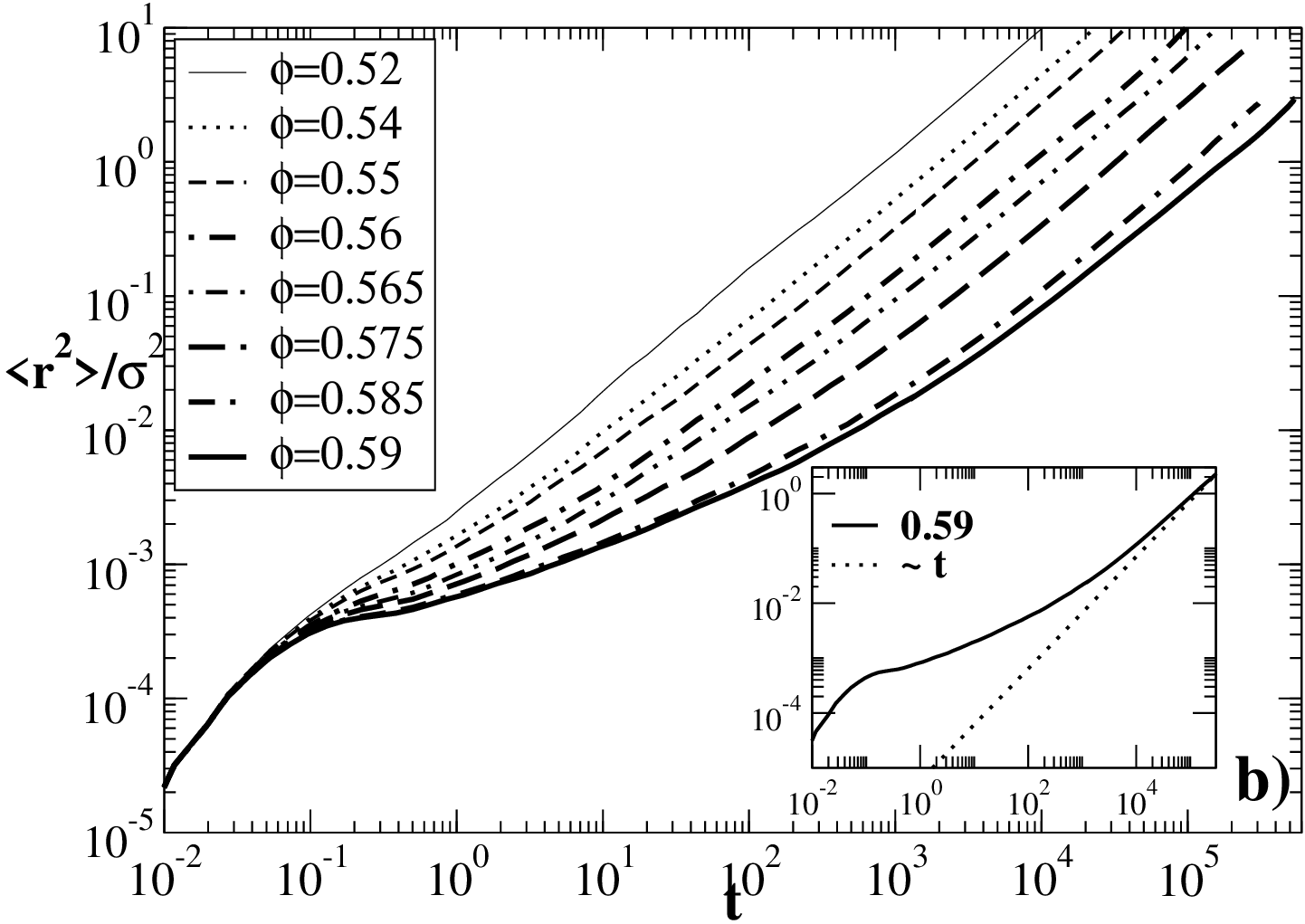,width=9.cm,clip=,angle=0.}
           }
\vspace*{0.5cm}
\caption{a) 
  As in Fig.\protect\ref{fig:corre-2}(a) for $T=0.3$. In the inset fits
  with streched exponential are shown for different value of $q$ at
  the same packing fraction $\phi=0.58$. \,\,\, b) As in
  Fig.\protect\ref{fig:corre-2}(b) for $T=0.3$. In the inset a fit as in
  Fig.\protect\ref{fig:corre-0.4}(b) is shown. There is no evident
  subdiffusive regime.}
\label{fig:corre-0.3}
\end{figure}

%%%%%LINEA

\begin{figure}
\centerline{\psfig{figure=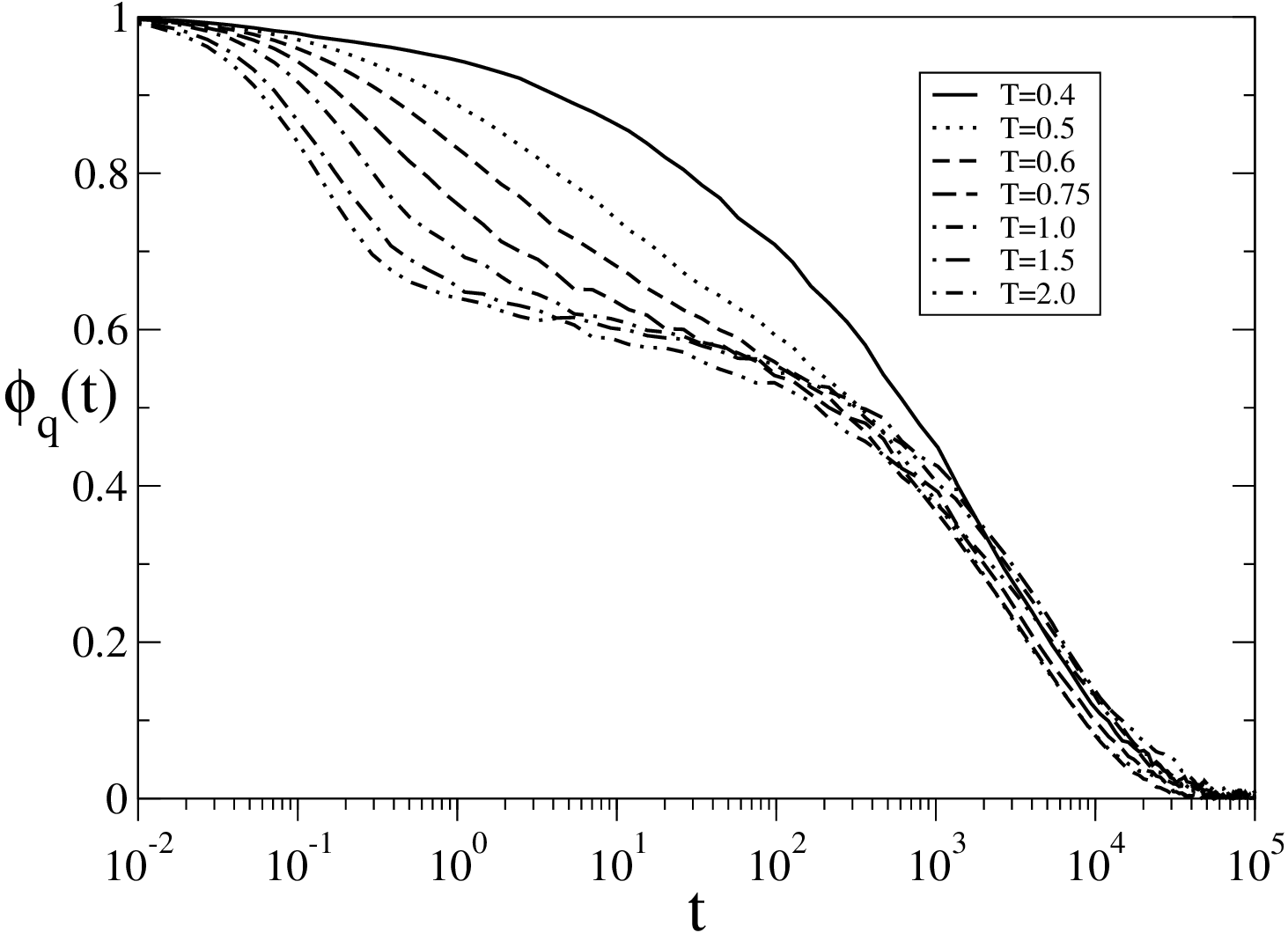,width=9.cm,clip=,angle=0.}
           }
\vspace*{0.5cm}
\caption{
  Density-density correlation function along iso-$D/D_0$ line
  $D=5\cdot 10^{-6}$.The wave-vector chosen corresponds for all cases
  to $q^*=2\pi/\sigma_A.$}
\label{fig:corre-linea}
\end{figure}

\begin{figure}
\centerline{\psfig{figure=fig13.eps,width=9.cm,clip=,angle=0.}
           }
\vspace*{0.5cm}
\caption{
  Mean square displacement along iso-$D/D_0$ line $D=5\cdot
  10^{-6}$.}
\label{fig:msd-linea}
\end{figure}

\begin{figure}
\centerline{\psfig{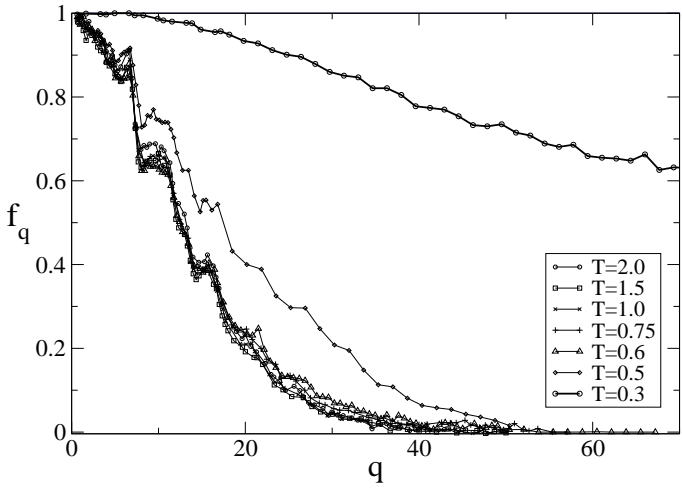}
           }
\vspace*{0.5cm}
\caption{
  Non ergodicity parameter $f_q$ along iso-$D/D_0$ line $D=5\cdot
  10^{-6}$ as obtained by fitting the correlation functions (see text
  for details).}
\label{fig:fq-linea}
\end{figure}

\begin{figure}\centerline{\psfig{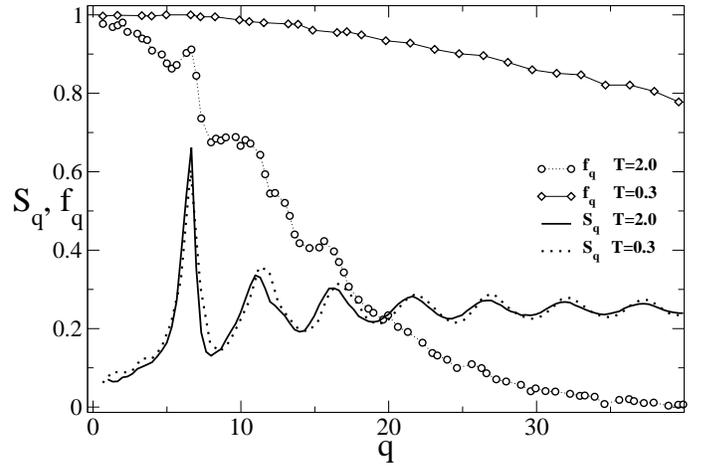}
           }
\vspace*{0.5cm}
\caption{Non ergodicity parameter $f_q$ and partial static
    structure factor $S_q$ at $T=2.0$ and $T=0.3$. The different shape
    of $f_q$ reflects the difference between attractive and repulsive
    glass (see text for details).}  \label{fig:fq-sq} 
\end{figure}

\end{multicols}


\begin{references}
\bibitem{nota0.3} 
        We note that this value is slightly higher
        than the one observed in Fig.~\ref{fig:corre-0.3}(b), because we
        are here looking at a point slightly less close to the glass
        transition, thus localization is a bit less effective for this
        case. However, when this happens, it is always inside the
        attractive well in this part of the phase diagram.
\bibitem{nota0.4} 
        Actually, using the stretched exponential, one can fit the
        correlators not too badly. However, the parameters that are
        found from the fits are completely unreliable. Indeed, for the
        amplitude parameter, which should give an estimate of the
        non-erogdicity parameter, we find for every $q$-vector, the
        value of 1. This is the reason why we do not consider these
        fits to have any validity.
\end{references}
\end{document}